\documentclass[prd,twocolumn,a4paper,superscriptaddress,floatfix]{revtex4-1}
\usepackage{graphicx}
\usepackage{amssymb}
\usepackage{amsfonts}
\usepackage{amsmath}
\usepackage{color}
\usepackage{ulem}
\usepackage[utf8]{inputenc}
\usepackage[T1]{fontenc}

\begin{document}

\title{$\gamma^* \gamma$ and $\gamma^* p$ scattering in IHQCD}

\author{Artur Amorim}
\affiliation{Centro de F\'{\i}sica do Porto e Departamento de F\'{\i}sica e Astronomia da Faculdade de Ci\^encias da Universidade do Porto, Rua do Campo Alegre 687, 4169-007 Porto, Portugal}
\author{Miguel S. Costa}
\affiliation{Centro de F\'{\i}sica do Porto e Departamento de F\'{\i}sica e Astronomia da Faculdade de Ci\^encias da Universidade do Porto, Rua do Campo Alegre 687, 4169-007 Porto, Portugal}

\begin{abstract}
We perform joint fits using measurements of the proton structure functions $F_2(x,Q^2)$ and $F_L(x,Q^2)$, of the photon structure function $F_2^\gamma(x,Q^2)$ and of the total cross-sections $\sigma\left( \gamma p \rightarrow X\right)$ and $\sigma \left( \gamma \gamma \rightarrow X\right)$. The data is gathered from several sources including HERA and LEP. The kinematical range considered is wide with a photon virtuality $Q^2 \leq400 \, {\rm GeV}^2$, Bjorken variable $x<0.01$ and $\sqrt{s} > 4 \, {\rm GeV}$. This is done by considering a $U\left(1\right)$ vector gauge field minimally coupled to the graviton Regge trajectory in improved holographic QCD. We find good agreement with the data for a joint fit of $F_2^\gamma$ and $\sigma\left(\gamma \gamma \rightarrow X\right)$ with a  $\chi_{d.o.f.}^2 = 1.07$. For a joint fit of 
$F_2$, $F_L$ and  $\sigma\left( \gamma p \rightarrow X\right)$ we find $\chi_{d.o.f.}^2 = 1.40$. A joint fit of all observables for both processes $\gamma^* \gamma$ and $\gamma^* p$ gives a $\chi_{d.o.f.}^2 = 1.38$.
The gravitational couplings of the $U\left(1\right)$ vector gauge field and of the proton with the Reggeons of the graviton Regge trajectory  are given as  an output of the fitting procedure. An approximate relation between the structure functions, valid at NLO QCD, is used to estimate the parton distribution function of the gluon inside the proton.
\end{abstract}
\maketitle

\section{Introduction}

Holography has been quite successful in describing QCD processes dominated by Pomeron exchange. The QCD Pomeron is dual to the graviton Regge trajectory \cite{brower_pomeron_2007}, 
hence modelling the external scattering states and the exchange of the graviton trajectory one can compare holographic predictions with experiment. This comparison is done in a kinematical window 
where the QCD interaction is dominated by a gluon rich medium, away form the perturbative QCD regime. This is the regime of small Bjorken variable $x$, also known as low $x$ QCD. In this context,
most cases explored in the literature \cite{BallonBayona:2007qr, hatta_deep_2008, cornalba_saturation_2008,pire_ads/qcd_2008,albacete_dis_2008,hatta_relating_2008, brower_saturation_2008, levin_glauber-gribov_2009, brower_elastic_2009, gao_polarized_2009, hatta_polarized_2009, kovchegov_comparing_2009, avsar_shockwaves_2009, domokos_pomeron_2009, cornalba_deep_2010, dominguez_particle_2010, cornalba_ads_2010, betemps_diffractive_2010, gao_polarized_2010, kovchegov_$r$_2010, levin_inelastic_2010, domokos_setting_2010, brower_string-gauge_2010, costa_deeply_2012, Brower:2012mk, stoffers_zahed_pomeron_2012, costa_vector_2013, anderson_central_2014, kovensky_struct_2014, kovensky_DIS_2015, Ballon-Bayona:2015wra, kovensky_DIS_2016, ballon_bayona_unity_2017, Nally:2017nsp, kovensky_DIS_2018, lee_ryu_zahed_vmp_2018, Amorim:2018yod, kovensky_F1_F2_2019, kiminad_zahed_gpd_2019, FolcoCapossoli:2020pks} include the scattering of an off-shell photon produced by an incoming electron with a target proton ($\gamma^*p$ scattering).
One process that has been less explored  is $\gamma^*\gamma^*$ scattering, which arises in high-energy 
$e^{+}e^{-}$ interactions for which some electrons and positrons are scattered by emitting virtual photons whose virtualities can be computed by measuring the angles and energies of the scattered electrons and positrons. The virtual photons can then fluctuate into quark-anti-quark pairs and hence generate a hadronic final state $X$. It is in this context that we refer to $\gamma^{*} \gamma^{*}$ scattering.

Currently the proton is seen as a collection of partons, each carrying a fraction $x$ of the longitudinal momentum. The probability density of finding a given parton carrying a momentum fraction $x$ at a squared energy scale $Q^2 = -q^2$ is known as a Parton Distribution Function (PDF). The precise knowledge of the PDFs is vital to test predictions of the Standard Model and beyond Standard Model models in the LHC. However, these objects are nonperturbative and cannot be derived from first principles in QCD. Instead the functional dependence of the PDFs on $x$ is parameterised at some high enough scale $Q^2 = Q_0^2$ where nonperturbative effects are not important. These input distributions can then be evaluated at another scale $Q^2$ using the DGLAP equations, provided the formalism of perturbative QCD is adequate. The input parameters are then fixed by data from different experiments. Despite the successes of this procedure, perturbative QCD techniques, like the BFKL pomeron
  \cite{Fadin:1975cb,Kuraev:1977fs,Balitsky:1978ic}, breakdown in the low $x$ kinematical regime where the gluons dominate.

The goal of this work is to extend previous work on the holographic description of Pomeron exchange to include the longitudinal proton structure function $F_L^p$, the total cross-section $\sigma\left(\gamma p \rightarrow X\right)$ as well a new class of $\gamma^{*} \gamma^{*}$ processes. The latter are a much cleaner application of holography because 
an off-shell photon generates a source for the quark bilinear operator $J^\mu = \bar{\psi} \gamma^\mu \psi$ which, according to holography, is dual to the non-normalizable mode of a bulk $U(1)$ gauge field.
Describing a proton as an external on-shell state is notoriously more difficult in holography. The Pomeron coupling to the current $J^\mu$ is holographically equivalent to the 
 interaction between the $U(1)$ vector gauge field and the graviton Regge trajectory.

We shall follow our previous works \cite{ballon_bayona_unity_2017, Amorim:2018yod} that use the Improved holographic QCD (IHQCD) 
model constructed in \cite{gursoy_exploring_2008, gursoy_exploring_2008-1, gursoy_improved_2011}. We start by fitting the product of the bulk couplings between the bulk fields dual to $J^\mu$ and the proton with the Reggeons of  the graviton Regge trajectory as well the parameters of the 
Pomeron kernel. This initial fit has nine parameters which are fixed by the HERA data of $F_2$ and $F_L$ presented in \cite{Aaron:2009aa,Collaboration:2010ry} as well $\sigma\left(\gamma p \rightarrow X\right)$ data from \cite{pdg_2018}.  The fit uses 358 data points, covering the very large kinematical range of $x<10^{-2}$ and $Q^2 \leq 400 \, {\rm GeV}^2$ for $F_2(x,Q^2)$ and  $Q^2 \leq 45 \, {\rm GeV}^2$ for $F_L(x,Q^2)$ and $\sqrt{s} > 4.6 \, {\rm GeV}$ for $\sigma\left(\gamma p \rightarrow X\right)$, where $Q^2$ is the photon virtuality. We have found a $\chi^2$ of 1.40.  We shall then use the fixed Pomeron kernel parameters and $F_2^\gamma(x,Q^2)$ and $\sigma\left(\gamma \gamma \rightarrow X\right)$ data to fit the local bulk couplings of the bulk $U(1)$ gauge field with the Reggeons of the graviton Regge trajectory. For this fit we have found a $\chi^2$ of 1.07.

We also consider global fits,  including both $\gamma^{*}p$ and $\gamma^{*}\gamma$ processes. As we shall see, allowing all the parameters of the model to vary does not move these parameters significantly from the values determined before, therefore further extending the  range of Pomeron exchange processes described by holography in a consistent manner.

The gluon PDF can be extracted from both structure functions $F_2$ and $F_L$. This was done in the context of holographic QCD in \cite{Watanabe:2019zny} using the BPST kernel for pomeron exchange and a hard-wall in $AdS_5$ to add confinement effects.  The results presented in  \cite{Watanabe:2019zny}  were obtained using a $\chi^2$  fit with data for $Q^2 \leq 10 \, {\rm GeV}^2$. Since we have a good description of both $F_2$ and $F_L$  in a wider photon virtuality range, we use the same method to compute the holographic gluon PDF. We compare our results  with the NLO results of the CTEQ and NNPDF PDF sets, finding good agreement 
throughout the larger kinematical region.

This paper is organized as follows. We first review how to obtain expressions for the DIS structure functions $F_2(x,Q^2)$ and $F_L(x,Q^2)$, the photon structure function $F_2(x, Q^2)$ and the total cross-sections $\sigma\left(\gamma p \rightarrow X\right)$ and $\sigma\left(\gamma \gamma \rightarrow X\right)$ in generic
 AdS/QCD models. Later we focus on the improved holographic QCD model of \cite{gursoy_exploring_2008,gursoy_exploring_2008-1,gursoy_improved_2011} 
 and we fit our model for Pomeron exchange to data from different processes and experiments. Finally we discuss the quality of our fits and then we extract  the gluon PDF from our holographic computation of $F_2$ and $F_L$.
%%%%%%%%%%%%%%%%%%%%%%%%%%%%%%%%%%%
\section{$\gamma^{*}p$  observables}
%%%%%%%%%%%%%%%%%%%%%%%%%%%%%%%%%%%

In this section we review how to compute the structure functions $F_2$ and $F_L$ and the total cross-section $\sigma\left( \gamma p \rightarrow X\right)$ holographically. Details can be found in~\cite{ballon_bayona_unity_2017} and hence we will just present the main formulas that allow us to derive expressions for $F_2$, $F_L$ and $\sigma\left( \gamma p \rightarrow X\right)$ in terms of holographic quantities.

Using the optical theorem the structure functions can be related to the amplitude for forward Compton scattering 
\begin{align}
	A^{FC}(q, P) = & i {\left(2 \pi\right)}^4 \delta^4 \Big( \sum_i k_i \Big) \left[ n_T^2 \tilde{F}_1(x, Q^2) + \right. \notag \\ 
	& \left. + \frac{2x}{Q^2} \left( n_T\cdot P \right)^2 \tilde{F}_2(x, Q^2)\right],
\label{eq:amplitude_FC}
\end{align}
where $q$ is the momentum of the incoming photon, $P$ is the momentum of the incoming hadron and $n_T=n_T(q)$ is the transverse projection of the virtual photon polarization $n^\mu$. The DIS structure functions are extracted from the forward Compton amplitude through
\begin{equation}
F_i (x, Q^2 ) = 2 \pi \, {\rm Im} \tilde{F_i} (x, Q^2 ),\ \ \ \ \ (i=1,2)\,
\label{eq:struct_from_compton}
\end{equation}
and $F_L=F_2-2xF_1$.

Before showing how to compute (\ref{eq:amplitude_FC}) holographically let us  introduce the kinematics. We use light-cone coordinates $\left(+,-,\perp \right)$, with the flat space metric given by $ds^2 = - dx^+ dx^- + d x^2_\perp$, where $x_\perp \in \mathbb{R}^2$ is a vector in impact parameter space. 
We take for the large $s$ kinematics of  $12\to34$ scattering the following
\begin{equation}
k_1=\left(\!\sqrt{s},-\frac{Q^2}{\sqrt{s}} ,0\right),
\ \ \ \ \ 
k_2=\left(\frac{M^2}{\sqrt{s}},\sqrt{s} ,0\right),
\label{eq:kinematics}
\end{equation}
where $k_1$ is the  incoming photon momenta and $k_2$ is the incoming proton 
momentum with mass $M$. For the forward Compton scattering amplitude the 
momentum transfer $q_\perp = 0$ so that the outgoing photon has $k_3=-k_1$ and the outgoing
proton $k_4=-k_2$. The  incoming and outgoing photon polarizations are the same.
The possible polarization vectors are
\begin{equation}
  \label{eq:polarization vectors} 
n(\lambda) = \begin{cases}
    \big(0,0,\epsilon_\lambda\big) \,, & \ \ \ \lambda=1,2 \,,\\
   \big( \sqrt{s}/Q, Q/\sqrt{s},0  \big)\, , & \ \ \ \lambda=3\,,
    \end{cases}
\end{equation}
where $\epsilon_\lambda$ is just the usual transverse polarization vector.

\begin{figure}[!t]
  \center
  \includegraphics[height=4cm]{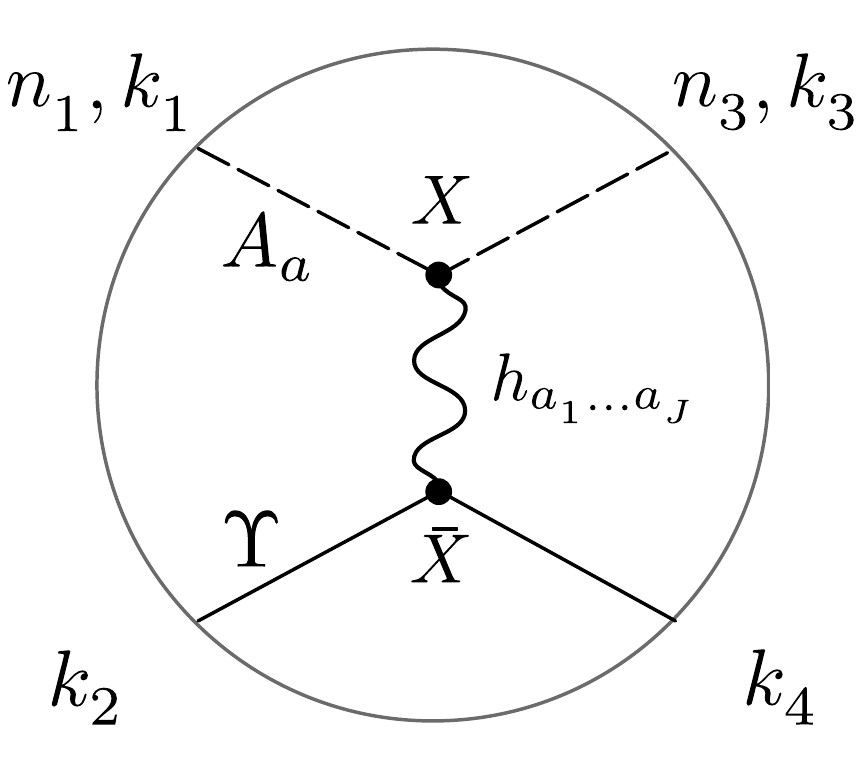} 
  \caption{Tree level Witten diagram representing spin $J$    exchange in  $12\to34$ scattering. 
The $n_1$ and $n_3$ labels denote the incoming/outgoing photon polarizations, for forward scattering $n_1=n_3$.
}
  \label{fig:Witten_diagram_gp}
\end{figure}

In the framework of AdS/QCD the above scattering amplitude  can be computed with the Witten diagram shown in figure~\ref{fig:Witten_diagram_gp}. The upper part of the diagram is related to the incoming and outgoing virtual photons, whereas the bottom part to the proton target. 
We are interested in the Regge limit where the amplitude is dominated by the exchange 
of the graviton Regge trajectory, which includes fields of even spin $J$. We also need to define our holographic external states. 
Among other fields, the holographic dual of QCD will have a dilaton field and a five-dimensional metric, which in the vacuum will have the form
\begin{equation}
ds^2 = e^{2 A(z)} \left[ dz^2 + \eta_{\mu \nu} dx^\mu dx^\nu \right]    \,, \qquad
\Phi = \Phi(z) \,, \label{IHQCDBackground}
\end{equation}
for some unknown functions $A(z)$ and $\Phi(z)$. 
We shall use greek indices in the boundary, with flat metric $ \eta_{\mu \nu}$.
We will work in the string frame.

In DIS the external photon is a source for the conserved $U(1)$ current $\bar{\psi} \gamma^\mu \psi$,
where the quark field $\psi$ is associated to the open string sector. 
The five dimensional dual of this current is a massless $U(1)$ gauge field $A$. We shall assume that this 
field is made out of open strings and that is minimally coupled to the metric, with the following action
\begin{equation}
\label{eq:u1 action} 
  S_A = - \frac{1}{4} \int d^5 X \sqrt{-g} \, e^{-\Phi} F_{ab}F^{ab}
\end{equation}
where $F=dA$ and we use the notation $X^a=(z,x^\alpha)$ for five-dimensional points. 
We will fix the gauge of the $U(1)$ bulk field to be $D_a A^a = 0$, which gives  
$A_z = 0$ and $\partial_\mu A^\mu = 0$. In this gauge, the solution to the equation of motion 
$ \nabla_a \left( e^{-\Phi} F^{ab} \right)=0$ is given by
\begin{equation}
 A_\mu ( X ; k,\lambda ) =  n_\mu(\lambda) \, f_Q ( z )\,e^{i k \cdot x}\,,
\end{equation}
where $f_Q(z)$ solves the differential equation
\begin{equation}
  \label{eq:u1 eom gauge fixed}
  \left[-Q^2+e^{\Phi-A}\partial_z\left(e^{A-\Phi}\partial_z \right) \right]f_Q(z)= 0 \,.
\end{equation}
The momentum  $k$ and the polarisation vector $n(\lambda)$, given  in  (\ref{eq:polarization vectors}), satisfy
\begin{equation}
  k^2 = Q^2 \,, \qquad 
  n_z = 0 \,, \qquad
   k \cdot n = 0 \,.
\end{equation} 
The UV boundary condition $f(0)=1$ gives the non-normalizable solution,
since the off-shell photon acts as a source for the quark bilinear current $\bar{\psi} \gamma^\mu \psi$.
Later it will be useful to use the identities
\begin{align}
&F_{\mu\nu}\left(X; k, n\right) = 2 i k_{[\mu}n_{\nu]} f_Q (z) e^{i k \cdot x}\,,\notag \\
&F_{z\nu} \left(X; k, n\right) = n_\nu \dot{f}_Q (z) e^{i k \cdot x}\,,
\label{eq:F_expression}
\end{align}
where $\dot{}=\frac{d\ }{dz}$.

For the proton target we consider a scalar field $\Upsilon$ that represents an unpolarised proton described by a normalizable mode of the form
\begin{equation}
\Upsilon(X;p )= \upsilon_m(z) \, e^{i p\cdot x} \,,
\label{eq:proton}
\end{equation}
where $p$ is the momentum and $m^2=-p^2$.
As explained in detail in~\cite{ballon_bayona_unity_2017},
the specific details of the function $\upsilon_m(z)$ will not be important because it will appear in an integral 
that can be absorbed in the coupling between the pomeron and the proton.

Next, to compute the Witten diagram of  figure~\ref{fig:Witten_diagram_gp}, we need to consider the interaction between the external scattering states and  a spin $J$ field in the graviton Regge trajectory.
The higher spin fields come from the closed string sector,  while the external fields come from the open sector. 
Their coupling is done by extending the minimal coupling between the graviton and the external states. 
This issue has been discussed in detail in~\cite{ballon_bayona_unity_2017}, thus we will just write the final result. 
We start by decomposing the spin $J$ field  $h_{a_1 \dots a_J}$  in $SO(1,3)$ irreducible representations. Then, in the Regge limit,
we are only interested in the TT components of this field $h_{\alpha_1 \dots \alpha_J}$, with
$\partial^{\nu}h_{\nu \alpha_2 \dots \alpha_J}=0$ and $h^\nu_{\ \nu\alpha_3 \dots \alpha_J}=0$. 
The coupling between the  $U(1)$ gauge field and the  TT components of the spin $J$ field has the form 
\begin{equation}
  \kappa_J  \int  d^5 X \sqrt{-g} \, e^{-\Phi }   h_{\alpha_1\dots \alpha_J} F^{\alpha_1 a} \partial^{\alpha_2} \dots \partial^{\alpha_{J-1}}F^{\alpha_J}_{\ \ \, a}\, .
\label{eq:gauge_field_spin_J_coupling}
\end{equation}
For the scalar field 
$\Upsilon$ we have 
\begin{equation}
\bar{\kappa}_J \int d^5 X \sqrt{-g} \, e^{-\Phi } h_{\alpha_1 \dots \alpha_J}  \Upsilon \partial^{\alpha_1} \dots \partial^{\alpha_J} \Upsilon  \, .
\label{eq:scalar_field_spin_J_coupling}
\end{equation}

Using the ingredients we have just introduced the contribution to the forward Compton scattering amplitude due to spin J exchange can be computed. Then  one needs to 
sum over the fields with of spin $J =2,4, \dots$ in the graviton Regge trajectory. This sum can be converted into an integral in the complex $J$-plane through a Sommerfeld-Watson transform. From the resulting expression  and using (\ref{eq:struct_from_compton}) one obtains for the DIS structure functions 
 $F_1$ and $F_2$ \cite{ballon_bayona_unity_2017}
\begin{align}
 x F_i(x, Q^2) = &4\pi Q^2\!\int \!dz d\bar{z} \, P^{(i)}_{13}(Q^2,z) P_{24}(P^2,\bar{z}) \times \notag \\
 & \times {\rm Im} \big[ \chi(s, t=0,z,\bar{z}) \big] \,,
\end{align}
where
\begin{align}
 P^{(1)}_{13}(Q^2,z) &= e^{ A(z) - \Phi(z)} f_Q^2\,, \\
 P^{(2)}_{13}(Q^2,z) &= e^{ A(z) - \Phi(z)} \left[ f_Q^2 + \frac{1}{Q^2} (\partial_z f_Q)^2 \right], \\
 P_{24}(P^2, \bar{z}) &= e^{3 A(\bar{z}) - \Phi(\bar{z})} \Upsilon^2 (P^2, \bar{z})\,,
\end{align}
and 
\begin{align}
\chi(s, t, z, \bar{z}) = \frac{\pi}{4}  \int & \frac{d J}{2 \pi i} \frac{S(z,\bar{z})^{J-1} \left(1-(-1)^J\right)}{\sin ( \pi J )} \times \notag \\
&\times \frac{k_J \bar{k}_J}{2^J} G_J (z, \bar{z}, t)\,,
\end{align}
with $S(z,\bar{z}) = s e^{-A(z)-A(\bar{z})}$. The eikonal phase $\chi(s, t, z, \bar{z}) $ results from the 
 $+\cdots+,-\cdots-$ component of the spin J propagator  $\Pi_{a_1\dots a_J, b_1 \dots b_J}\left(X, \bar{X}\right)$, which satisfies the identity
 \begin{align}
&\int \frac{dw^+ dw^- d^2 l_\perp}{2} \,  e^{-i q_\perp \cdot l_\perp} \, \Pi_{+\dots+, - \dots -}\left(X, \bar{X}\right) = \notag \\
& =  - \frac{i}{2^J} {\left( e ^{A+\bar{A}}\right)}^{J-1} G_J \left(z, \bar{z}, t\right) 
 \label{eq:propagator_identity}
\end{align}
where $w=x-\bar{x}=\left(w^+, w^-, l_\perp\right)$. $G_J ( z, \bar{z}, t )$ admits the spectral representation 
\begin{align}
    G_J ( z, \bar{z}, t ) = e^{B(z) + B(\bar{z})} \sum_n \frac{\psi_n (J, z ) \, \psi^{*}_n (J, \bar{z})}{t_n( J ) - t}\,,
    \label{eq:TransProp}
\end{align}
where $\psi_n (J, z )$ are the normalizable modes associated to the spin $J$ fields, that is they describe massive spin $J$ glueballs.
The function $B(z)$ depends on the particular holographic QCD model, for the model here considered $B=\Phi - A/2$.

The next step is to assume that the $J$-plane integral can be deformed from the poles at even J, to the poles $J = j_n (t)$ defined by $t_n (J) = t$. 
The scattering domain of negative $t$ contains these poles along the real axis for $J < 2$. After this step the expressions for $F_1$ and $F_2$ become
\begin{align}
&2 x F_1 (x, Q^2) = \sum_n {\rm Im}\,g_n x^{1-j_n(0)} Q^{2 j_n(0)} \bar{P}_{13}^{(1,n)} (Q^2)\,, \\ 
&F_2(x, Q^2) = \sum_n {\rm Im }\,g_n x^{1-j_n(0)} Q^{2 j_n(0)} \bar{P}_{13}^{(2,n)}(Q^2)\,,
\label{eq:holographic_structure_functions}
\end{align}
where
\begin{align}
&\bar{P}_{13}^{(1,n)} = \int dz \,  f_Q^2\, e^{(2-j_n)A+B- \Phi}  \psi_n(j_n ,z), \\
&\bar{P}_{13}^{(2,n)} = \int dz \left[ f_Q^2 + \frac{1}{Q^2} (\partial_z f_Q)^2 \right] e^{(2-j_n)A+B-\Phi}\psi_n(j_n,z),
\end{align}
with $j_n$ evaluated at $t = 0$. The constants $g_n$, which involve the AdS local couplings and an integral over the proton wavefunction,  will be used as fitting constants of the model.  They are defined by
\begin{align}
g_n = & - \frac{\pi}{2} \left( i + \cot \frac{\pi j_n}{2} \right) \frac{\kappa_{j_n} \bar{\kappa}_{j_n}}{2^{j_n}} \frac{d j_n}{dt} \times \\
& \times \int d\bar{z} e^{-\left(j_n - 7/2 \right)A} \upsilon_m^2(\bar{z}) {\psi}_n^*(\bar{z})\,.
\label{eq:gn_def_gammap}
\end{align} 

The above discussion is aplicable to any holographical model of QCD. We shall consider the improved holographic QCD model introduced in 
\cite{gursoy_exploring_2008,gursoy_exploring_2008-1,gursoy_improved_2011}. Solving the model such that the spectrum of the scalar and tensor glueballs is reproduced fixes the
background fields  $A(z)$ and $\Phi(z)$, which give an approximate dual description of the QCD vacuum. We may then compute the non-normalizable modes for any $Q^2$ by solving numerically the equation (\ref{eq:u1 eom gauge fixed}) with the UV boundary condition $f_Q(0) = 1$.

All that is left is the equation of motion for the spin $J$ fields that are dual to the twist two operators, whose exchange gives the dominant 
contribution in DIS at low $x$. This equation is then analytically continued in $J$, in order to do the  Sommerfeld-Watson transform in Regge theory. 
As described in detail in \cite{ballon_bayona_unity_2017} the normalisable modes of the spin $J$ field $\psi_n(z)$ solve a Schr\"{o}dinger problem
\begin{equation}
  \left(-\frac{d^2}{dz^2}+U_J(z)\right)\psi_n(z)=t_n\psi_n(z)\,,
\end{equation}
where
\begin{align}
 & U_J(z)= \frac{3}{2}\left(\ddot A - \frac{2}{3}\ddot \Phi\right) + \frac{9}{4}{\left(\dot A - \frac{2}{3}\dot \Phi \right)}^2 + \\
 &+ (J - 2)e^{-2A}\bigg[\frac{2}{l^2_s}\left(1+\frac{d}{\sqrt{\lambda}}\right) + \frac{J + 2}{\lambda^{4/3}} + \bigg. \notag \\
 & \bigg. + e^{2A} \left(a \ddot \Phi + b \left(\ddot A - \dot{A}^2\right) + c \dot{\Phi}^2  \right)\bigg]\,.
 \nonumber
\end{align}
The  first line in this equation  represents the potential for the graviton and the remaining terms deform the graviton potential.
This potential is analytically continued in $J$ in such a way that  the value of the intercept $J=j_n$ 
is obtained when the $n$-th  eigenvalue satisfies $t_n(J)=0$. We will use 
a Chebyshev algorithm with 1000 points to compute the eigenvalues $t_n$ and the eigenfunctions $\psi_n$.
From the low energy effective string theory perspective, $l_s$ is related to the string tension; $d$ is related to the anomalous dimension curve of the twist 2 operators, or it can also be thought as encoding the information of how the masses of the closed strings excitations are corrected in a slightly curved background; the constants $a$, $b$ and $c$ encode the first order derivative expansion in effective field theory.
All these  constants will be adjusted from  fitting  $F_2$, $F_L$ and $\sigma\left( \gamma p \rightarrow X\right)$  data. 

Finally, the DIS structure functions $F_2$ and $F_L = F_2 - 2 x F_1$ can be written in Regge theory in the following form
\begin{align}
&F_2^p(x, Q^2) = \sum_{n} \frac{ {\rm Im} \, g_n}{4 \pi^2 \alpha} \, x^{1-j_n}  f_n^2 (Q^2) \,, \\
&F_L^p(x, Q^2) = \sum_{n} \frac{{\rm Im} \, g_n}{4 \pi^2 \alpha} \, x^{1-j_n}  f_n^L (Q^2) \, ,
\end{align}
where
\begin{align}
&f_n^2(Q^2) = Q^{2 j_n} \int dz \,e^{-\left(j_n-\frac{3}{2}\right)A}  \left( f_Q^2  +  \frac{\dot{f}_Q^{2}}{Q^2}      \right) \psi_n(z) \,,
\\
&f_n^L(Q^2) = Q^{2 j_n} \int dz \,e^{-\left(j_n-\frac{3}{2}\right)A}  \frac{\dot{f}_Q^{2}}{Q^2} \, \psi_n(z)\,.
\end{align}

The structure functions $F_2$ and $F_L$ are all related to the total cross-sections $\sigma^T_{\gamma^* p}$ and $\sigma^L_{\gamma^{*} p}$ of the inelastic process $\gamma^{*} p \rightarrow X$. Here T and L refer to the transverse and longitudinal polarisation of the incoming off-shell photon. At low-$x$ the proton structure functions are given by
\begin{align}
&F_2^p\big(x, Q^2\big) = \frac{Q^2}{4 \pi^2 \alpha} \Big[ \sigma^T_{\gamma^* p}\big(s, Q^2\big) +  \sigma^L_{\gamma^* p}\big(s, Q^2\big)   \Big] \,,\\
&F_L^p\big(x, Q^2\big) = \frac{Q^2}{4 \pi^2 \alpha}\,  \sigma^L_{\gamma^* p}\big(s, Q^2\big)\,,
\end{align}
and since only on-shell photons have transverse polarisation, the total cross-section of the process $\gamma p \rightarrow X$ is related to the proton structure function $F_2^p\left(x, Q^2\right)$ through
\begin{equation}
\sigma(\gamma p \rightarrow X) = 4 \pi^2 \alpha \lim_{Q^2 \rightarrow 0} \frac{F_2^p \big(x, Q^2\big)}{Q^2}\,.
\label{eq:total_sigma}
\end{equation}
Taking the limit (\ref{eq:total_sigma}) we  obtain the  holographic expression for the total  cross-section
\begin{equation}
\sigma(\gamma p \rightarrow X) =  \sum_{n} {\rm Im} \, g_n \, s^{j_n -1 } \int dz \,e^{-\left(j_n-\frac{3}{2}\right)A}  \psi_n(z)\,.
\end{equation}

%%%%%%%%%%%%%%%%%%%%%%%%%%%%%%%
\section{$\gamma^{*}\gamma^*$ observables}
\label{sec:gamma_gamma}
%%%%%%%%%%%%%%%%%%%%%%%%%%%%%%%

In this section we describe the observables in $\gamma^{*}\gamma^*$  scattering and how to compute them using holography. In this process photons can reveal either their point-like or  hadron-like behaviour.
In the point-like case one of the quarks takes part in the hard interaction while in the hadron-like case the photons fluctuate into hadrons with the same quantum numbers of the photon (i.e. vector mesons like $\rho$, $\omega$, $\phi$) and the interaction is the same as in hadron-hadron scattering. These pictures coexist together and are dominant in different kinematical regions. For high transverse momentum of the quarks or high virtuality of one of the photons the point-like nature is dominant. For lower values of the photon virtuality the interaction is spread over longer times, giving time for the quarks to form bound states through gluon exchange. This is the regime we are interested here.

The process $e^{+}e^{-} \rightarrow e^{+}e^{-} X$ is factorised into two three terms: one for the radiation of  the virtual photon from the electron, one for the radiation of the other virtual photon from the positron and the term that couples the $\gamma^{*} \gamma^{*}$ system to the final hadronic state $X$. Like in $e^-p$ DIS we can define the following variables
\begin{align}
&y_1 = \frac{k_1 \cdot p_2}{p_1 \cdot p_2}\,, \quad y_2 = \frac{k_2 \cdot p_1}{p_1 \cdot p_2} \,,\\
&x_1 = \frac{Q_1^2}{2 k_1 \cdot p_2}\,, \quad x_2 = \frac{Q^2_2}{2 k_2 \cdot p_1}\,, \notag
\end{align}
where $p_1$ and $p_2$ are the incoming leptons momenta, and
$k_1$ and $k_2$ the momenta of the two off-shell photons with virtualities $Q_1^2 = k_1^2$ and $Q_2^2 = k_2^2$.
 In terms of these variables we can write the differential cross-section for $e^{+}e^{-} \rightarrow e^{+}e^{-} X$ as~\cite{donnachie_dosch_landshoff_nachtmann_2002}
\begin{align}
\label{eq:ee_eeX_sigma}
&\frac{d^4 \sigma}{d y_1 d y_2 d Q_1^2 d Q_2^2} = {\left( \frac{\alpha}{2 \pi} \right)}^2 \times \notag \\
&\times \left[ P^{T}_{\gamma/e^{-}} \big(y_1, Q_1^2\big) P^{T}_{\gamma/e^{+}} \big(y_2, Q_2^2\big) \sigma^{TT} \big(Q_1^2, Q_2^2, W^2\big) \right. \notag \\
&\left. + P^{T}_{\gamma/e^{-}} \big(y_1, Q_1^2\big) P^{L}_{\gamma/e^{+}} \big(y_2\big) \sigma^{TL} \big(Q_1^2, Q_2^2, W^2\big) \right. \notag \\
&\left. + P^{L}_{\gamma/e^{-}} \big(y_1\big) P^{T}_{\gamma/e^{+}} \big(y_2, Q_2^2\big) \sigma^{LT} \big(Q_1^2, Q_2^2, W^2\big) \right. \\
&\left.  + P^{L}_{\gamma/e^{-}} \big(y_1\big) P^{L}_{\gamma/e^{+}} \big(y_2\big) \sigma^{LL} \big(Q_1^2, Q_2^2, W^2\big) \right] \frac{1}{Q_1^2 Q_2^2}\,,
\notag
\end{align}
where the functions $P^T_{\gamma / e}$ and $P^L_{\gamma / e}$ are related to flux factors and are given by
\begin{align}
& P^{T}_{\gamma/e} \big(y, Q^2\big) = \frac{1+{\left(1-y\right)}^2}{y} - \frac{2 m_e^2 y}{Q^2}\,, \\
& P^{L}_{\gamma/e} (y) = 2\frac{1-y}{y}\,.
\end{align}
The cross sections $\sigma^{ij}(Q_1^2, Q_2^2, W^2)$, with $i,j=T,L$, are the total cross sections for $\gamma^*(Q_1^2) \gamma^{*}(Q_2^2) \rightarrow {\rm X}$ for 
incoming photons with transverse (T) or longitudinal (L) polarization in their centre-of-mass frame. 
The differential cross-section expression also results from integrating over the angle between the plane of the scattered leptons in the center-of-mass frame of the $\gamma^* \gamma^*$ system and does not include terms that are present for polarized lepton beams.

\begin{figure}[!h]
  \center
  \includegraphics[height=4cm]{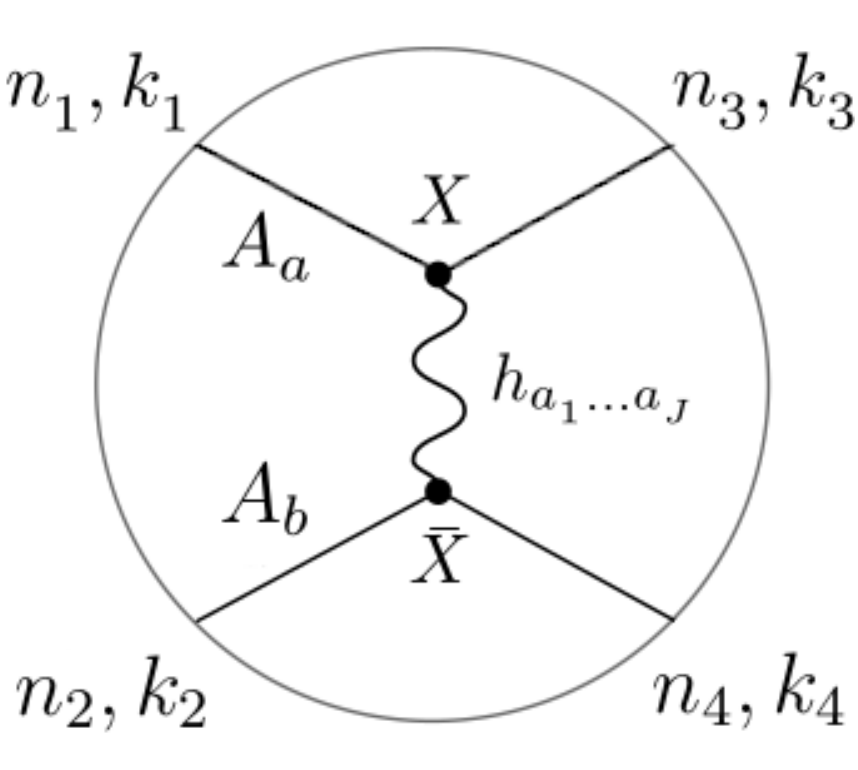} 
  \caption{Tree level Witten diagram representing spin $J$    exchange in a $2\to2$ scattering process between bulk photons. 
  We will consider a kinematical regime dominated by the exchange of spin $J$ fields in the graviton Regge trajectory.}
  \label{fig:Witten_diagram_gamma_gamma}
\end{figure}

The observables we want to study can be expressed in terms of the cross-sections $\sigma^{ij}(Q_1^2, Q_2^2, W^2)$. Hence we will compute these cross-sections holographically starting from the general scattering amplitude of 
$\gamma^{*}(Q_1)\gamma^{*}(Q_2) \rightarrow \gamma^{*}(Q_3) \gamma^{*}(Q_4)$ for arbitrary polarisation of the incoming and outgoing virtual photons. Later, from this general amplitude, we specify for the case of the cross-section $\sigma(\gamma \gamma \rightarrow X)$ and of the structure function $F_2^\gamma(x, Q^2)$. 
The Witten diagram that is relevant for our calculations is the one in figure~\ref{fig:Witten_diagram_gamma_gamma}. The off-shell photons source the quark-bilinear operator current $\bar{\psi}\gamma^\mu\psi$ so the 
$\gamma^{*}\gamma^{*} \rightarrow \gamma^{*}\gamma^{*}$ amplitude can be computed form the  $2\rightarrow 2$  scattering of  bulk photons described by a  $U(1)$ gauge field.

Finally, we describe the kinematics we will be considering. 
Using the same light-cone coordinates as in the previous section, 
the incoming photons have the space-like momenta
\begin{equation} 
k_1 = \left( \sqrt{s}, - \frac{Q_1^2}{\sqrt{s}}, 0 \right) \,,  \qquad  
k_2 = \left( - \frac{Q_2^2}{\sqrt{s}}, \sqrt{s},  0 \right) \,,
\end{equation}
as well as the outgoing photons,
\begin{align}
&k_3 = - \left( \sqrt{s},  \frac{q_\perp^2 - Q_3^2}{\sqrt{s}}, q_\perp \right) , \notag \\ 
&k_4 = - \left(  \frac{q_\perp^2-Q_4^2}{\sqrt{s}}, \sqrt{s},  -q_\perp \right),
\end{align}
where $Q^2_i = k_i^2 >0$  $(i=1, \dots, 4)$ is the off-shellness. 
We consider the Regge limit of large s and fixed $t = -q_\perp^2$.
We can now define the photons polarization vectors $n^\lambda_i$ that satisfy the condition
$n_i^\lambda \cdot k_i = 0$. 
The incoming off-shell photons have the following polarization vectors
\begin{align}
    &n_1=
    \begin{cases}
      \left(0,0,1,0\right), & \lambda=1 \\
      \left(0,0,0,1\right), & \lambda=2 \\
      \frac{1}{Q_1} \left( \sqrt{s}, \frac{Q_1^2}{\sqrt{s}}, 0, 0 \right), & \lambda = 3
    \end{cases}\,,\\
    &n_2=
    \begin{cases}
      \left(0,0,1,0\right), & \lambda=1 \\
      \left(0,0,0,1\right), & \lambda=2 \\
      \frac{1}{Q_2} \left( \frac{Q_2^2}{\sqrt{s}}, \sqrt{s}, 0, 0 \right), & \lambda = 3
    \end{cases} \,,
    \label{eq:inPolarization}
\end{align}
while for the outgoing photons we have 
\begin{align}
    &n_3=
    \begin{cases}
      \left(0,\frac{2 q_x}{\sqrt{s}},1,0\right), & \!\lambda=1 \\
      \left(0,\frac{2 q_y}{\sqrt{s}},0,1\right), & \!\lambda=2 \\
      \frac{1}{Q_3} \left(\sqrt{s}, \frac{Q_3^2+q_\perp^2}{\sqrt{s}}, q_\perp \right), & \!\lambda = 3
    \end{cases}\,, \\
    &n_4=
    \begin{cases}
      \left(-\frac{2 q_x}{\sqrt{s}},0,1,0\right), & \!\lambda=1 \\
      \left(-\frac{2 q_y}{\sqrt{s}},0,0,1\right), & \!\lambda=2 \\
      \frac{1}{Q_4} \left(\frac{Q_4^2+q_\perp^2}{\sqrt{s}},\sqrt{s},- q_\perp \right), & \!\lambda = 3
    \end{cases} \, .
    \label{eq:outPolarization}
\end{align}  
Notice that the transverse photons~$\left(\lambda=1,2\right)$ are normalized such that~$n^2 = 1$, while the longitudinal photons~$\left(\lambda=3\right)$ are normalized such that~$n^2 = -1$.

%%%%%%%%%%%%%%%%%%%%%%%%%%%%%%
\subsection{Computation of Witten diagram}
%%%%%%%%%%%%%%%%%%%%%%%%%%%%%%

The amplitude associated to the exchange of a spin J field in the Witten diagram of figure \ref{fig:Witten_diagram_gamma_gamma} is given by
\begin{align}
&A_J \left(k_i, \lambda_j\right) = - \kappa_J^2 \int d^5X d^5 \bar{X} \sqrt{-g} \sqrt{-\bar{g}} \,e^{-\Phi} e^{-\bar{\Phi}} \times \notag \\ 
&\times F_{- a}^{(1)} (k_1, z) \partial_{-}^{J-2}F^{a \, (3)}_{-} (k_3, z)  \times  \\
&\times F_{+ b}^{(2)} (k_2, \bar{z}) \bar{\partial}_{+}^{J-2}F^{b \, (4)}_{+} (k_4, \bar{z} )  
\,\Pi^{- \dots - , + \dots +}\left(X, \bar{X}\right).
\notag
\end{align}
Using the kinematics introduced above  we obtain
\begin{align}
&A_J \left(k_i, \lambda_j\right) =- \kappa_J^2 {\left( \frac{i}{2} \sqrt{s}\right)}^{2J-4} \int d^5X d^5\bar{X} \frac{e^{5 (A+\bar{A}) }}{4} \times \notag \\
&\times e^{-\Phi - \bar{\Phi}} {\left(4 e^{-2(A+\bar{A})}\right)}^J F_{-a}^{(1)}F^{a \, (3)}_{-} F_{+ b}^{(2)}F^{b \, (4)}_{+} \times \notag \\
& \times e^{-i q_\perp \cdot \left(x_\perp - \bar{x}_\perp\right)}  \Pi_{+\dots+, - \dots -}\left(X, \bar{X}\right) .
\end{align}
By performing the change of variables $w = x - \bar{x}$ and using the propagator identity in equation (\ref{eq:propagator_identity}) we arrive at the result
\begin{align}
&A_J \left(k_i, \lambda_j\right) =\frac{i V \kappa_J^2}{2^J} s^J \int dz d\bar{z}
\, e^{3\left(A+\bar{A}\right)} e^{- \Phi - \bar{\Phi}} \times \notag \\
& \times e^{-(1+J)(A+\bar{A})} F(1,3)F(2,4) G_J (z, \bar{z}, t)\,, 
\end{align}
where
\begin{equation}
F(i,j) = 
\begin{cases}
f_{Q_i} f_{Q_j} \,, & \lambda_i = \lambda_j = 1,2 \\
\dot{f_{Q_i}} \dot{f_{Q_j}} /(Q_iQ_j) \,, & \lambda_i = \lambda_j = 3
\end{cases}\,,
\end{equation}
and $G_J (z, \bar{z}, t)$ is given by (\ref{eq:TransProp}).

Next need to sum the above amplitude over the even spin-$J$ fields with $J>2$, as we briefly describer in the previous section. Such sum can be computed through a Sommerfeld-Watson transform
\begin{equation}
\frac{1}{2} \sum_J s^J + (-s)^J  \to -\frac{\pi}{2} \int \frac{dJ}{2 \pi i} \frac{s^J + (-s)^J}{\sin \pi J}\,.
\end{equation}
This assumes that an analytic continuation of the amplitude to the complex $J$-plane is possible. We now deform the integral from the poles at even $J$, to the poles $J = j_n\left(t\right)$ defined by $t_n(J) = t$. The scattering domain of negative $t$ contains these poles along the real axis for $J<2$. Thus the forward scattering amplitude ($t= 0$) of $\gamma^{*}(Q_1)\gamma^{*}(Q_2) \rightarrow \gamma^{*}(Q_3) \gamma^{*}(Q_4)$ is 
\begin{align}
&\mathcal{A}^{\lambda_1, \lambda_2, \lambda_3, \lambda_4}_{Q_1, Q_2, Q_3, Q_4}\left(s,t = 0\right) = - \frac{\pi}{2} \sum_n s^{j_n} \left[ i + \cot\left(\frac{\pi j_n}{2}\right) \right] \times \notag \\
& \times \frac{\kappa_{j_n}^2}{2^{j_n}} \frac{d j_n}{dt} \int dz e^{-\left(j_n - 3/2\right) A} F(1,3) \psi_n(z) \times \notag \\
& \times  \int d\bar{z} e^{-(j_n-3/2)\bar{A}} F(2,4) {\psi_n^*(\bar{z})}\,.
\label{eq:scattering_amplitude}
\end{align}

Finally, we remark that the total cross-sections $\sigma^{ij}(Q_1^2, Q_2^2, W^2)$, with $i,j=T,L$, can now be computed using the optical theorem and appropriate photon polarisations, as well by setting $Q_3^2 = Q_1^2$ and $Q_4^2 = Q_2^2$.

\subsection{Photon structure function $F_2^\gamma$}

We consider first scattering between a virtual  and an on-shell photon. In analogy with deep inelastic $e^{\pm} p$ scattering, we can think of this process as deep inelastic $e^{\pm} \gamma$ scattering. Just as for $e^{\pm} p$ scattering, we can define a hadronic tensor $W^{\mu \nu}$ and two structure functions related by
\begin{align}
& \frac{W^{\mu \nu}\left(x, Q^2\right)}{8 \pi^2 \alpha} = - \left( g^{\mu \nu} + \frac{q_1^\mu q_1^\nu}{Q^2}\right) F_1^\gamma\big(x, Q^2\big) + \notag \\
& + \frac{1}{q_1 \cdot q_2} \left( q_2^\mu + q_1^\mu \frac{q_1 \cdot q_2}{Q^2} \right) \left( q_2^\nu + q_1^\nu \frac{q_1 \cdot q_2}{Q^2} \right) F_2^\gamma \big(x, Q^2\big)\,,
\end{align}
such that the cross section for the process $e \gamma \rightarrow e X$ can be written as
\begin{equation}
\frac{d^2 \sigma}{dx dy} = \frac{4 \pi \alpha ^2}{x y Q^2} \left[ \left(1-y\right) F_2^\gamma(x, Q^2) + x y^2 F^\gamma_1(x, Q^2) \right]\,.
\end{equation}
The structure functions are related to the total cross-sections of equation (\ref{eq:ee_eeX_sigma}) through the relations
\begin{align}
	\label{eq:F2_def}
	&F_2^{\gamma} (x, Q^2)= \frac{Q^2}{4 \pi^2 \alpha} \left[ \sigma_{TT}(s, Q^2, 0) +   \sigma_{LT}(s, Q^2, 0)\right] , \\
	&2 xF_1^{\gamma} (x, Q^2)= \frac{Q^2}{4 \pi^2 \alpha} \,\sigma_{TT}\big(s, Q^2, 0\big)\,.
\end{align}

Before continuing with the holographic computation let us discuss in which kinematical region will it be applicable. The hadronic photon structure function $F_2^\gamma$ differs from the proton structure function due to the point-like coupling of the photon to the quarks. This coupling makes the photon structure function to rise towards large values of Bjorken $x$, while in the case of the proton it decreases. Moreover, $F_2^\gamma$ has positive scaling violations for all values of $x$, while $F_2^p$ has positive scaling violations only at small values of $x$. Also, the point-like part can be evaluated at all orders in perturbative QCD and dominates for large values of $Q^2$ and for values of $x > 0.1$.
On the other hand, the hadronic-like part can not be computed in perturbative QCD. Like its $F_2^p$ counterpart, only its evolution with $Q^2$ can be determined. As in the case of the proton, an ansatz for the $x$ dependence at some scale $Q_0^2$ is given as input to QCD evolution equations. This ansatz can be derived from the Vector Meson Dominance model, since the photon can fluctuate in a vector meson like the $\rho$ meson. After that one assumes that the $F_2^\rho$ structure function is the same as the $F_2^{\pi^0}$ structure function which has been measured experimentally. Then this hadron-like component can be evolved using perturbative QCD and we can compare it with the $F_2^\gamma$ structure function that contains both the point-like and hadron-like contributions. The result is that although the hadron-like component is not important for high values of $Q^2$ and $x > 0.1$, it clearly dominates for very small values of $x$, meaning that we can use the Pomeron exchange picture to study this process. Hence our holographic expression is only valid for $x < 0.01$. Thus we proceed to the holographic computation of $F_2^\gamma$ in this kinematical region.

As mentioned, one photon in this process is quasi-real. Let us represent such photon by the lower part of the Witten diagram. Then the $\bar{z}$ integral simplifies to 
\begin{equation}
\int d\bar{z} e^{- \left(j_n - 3/2\right) A} {\psi_n^* \big(j_n, \bar{z}\big)}\,.
\end{equation}
Defining
\begin{equation}
g_n = - \frac{\pi}{2} \left( i + \cot \frac{\pi j_n}{2} \right) \frac{\kappa^2_{j_n}}{2^{j_n}} j_n' \int d\bar{z} e^{- \left(j_n - 3/2\right) A} {\psi_n^* \big(j_n, \bar{z}\big)}\,,
\label{eq:gn_def_gammagamma}
\end{equation}
and using the optical theorem, we can write the total cross-sections $\sigma_{TT}$ and $\sigma_{LT}$ as
\begin{align}
&\sigma_{TT} =  \sum_n {\rm Im} \, g_n s^{j_n - 1} \int dz e^{-\left(j_n - 3/2\right) A} f_Q^2 \,\psi_n(z) \,, \\
&\sigma_{LT} =  \sum_n  {\rm Im} \, g_n s^{j_n - 1} \int dz e^{-\left(j_n - 3/2\right) A} \frac{\dot{f}_Q^2}{Q^2} \,\psi_n(z)\,.
\end{align}
Using (\ref{eq:F2_def}) and $s  = Q^2 / x$, the holographic expression for $F_2^\gamma$ is then
\begin{align}
& F_2^\gamma\big(x, Q^2\big) = \sum_n \frac{ {\rm Im}  \, g_n}{4 \pi^2 \alpha} \, x^{1- j_n} f_n^{\gamma} (Q^2) \, \\
& f_n^{\gamma} (Q^2) = Q^{2 j_n}  \int dz e^{- \left( j_n - 3/2\right) A} \left( f_Q^2 + \frac{\dot{f}_Q}{Q^2} \right) \psi_n(z) \,.
\end{align}

%%%%%%%%%%%%%%%%%%%%%%%%%%%%%%%%%
\subsection{Total cross section $\gamma \gamma \rightarrow X$}

In this process both photons are considered quasi-real, i.e. $Q_1^2 \approx 0$ and $Q_2^2 \approx 0$. In our holographic setup the non-normalizable modes of the bulk $U\left(1\right)$ gauge field
satisfy
\begin{equation}
\lim_{Q \rightarrow 0} f_Q \left(z\right) = 1\,, \quad \lim_{Q \rightarrow 0} \frac{\dot{f_Q}}{Q} = 0\,.
\end{equation}
Therefore, using   (\ref{eq:scattering_amplitude}) and the optical theorem,  the cross sections 
$\sigma^{LL}( 0, 0, W)$, $\sigma^{TL}(0, 0, W)$ and  $\sigma^{LT}(0, 0, W)$ vanish. This is expected since real photons only have transverse polarisation and hence the cross-sections that involve at least one longitudinal on-shell photon do not contribute to this process.
Thus
\begin{align}
\sigma(\gamma \gamma &\rightarrow X) =  \sigma^{TT}\big(0, 0, s = W^2\big) = \notag \\
&= \sum_n {\rm Im} \, g_n s^{j_n - 1} \int dz e^{- \left( j_n - 3/2\right) A}  \psi_n(z)\,.
\end{align}
Like in the case of $\gamma^{*}p$ processes the numbers ${\rm Im} \, g_n$ have the same definition as the ones in our  holographic expression $F_2^\gamma$ and these observables are related by
\begin{equation}
\sigma(\gamma \gamma \rightarrow X) = 4 \pi^2 \alpha \lim_{Q^2 \rightarrow 0} \frac{F_2^\gamma}{Q^2}\,.
\end{equation}

\section{Data analysis and results}

With the previously described setup we proceed to find the best values for the potential parameters $l_s$, $a$, $b$, $c$ and $d$, as well as for the coupling values ${\rm Im} \, g_n$. The ${\rm Im} \, g_n$ constants have different definitions for $\gamma^*\gamma$ and $\gamma^*p$ processes, so we will determine a set of values  for each process class.
We use the first four Reggeons, which are enough to reproduce the non-trivial $x$ behaviour of the proton structure functions $F_2^p$, $F_L^p$ and the total cross-section $\sigma\left(\gamma p \rightarrow X \right)$.  Adding several trajectories explains the so-called  hard-pomeron behaviour  for large $Q$ and the soft-pomeron 
 behaviour for smaller $Q$, as discussed  in \cite{ballon_bayona_unity_2017}. Each ${\rm Im} \, g_n$ is associated with a Reggeon.

We find the best set of parameter values $\alpha_i$ by minimising the $\chi^2$ quantity
\begin{align}
\chi^2 = \sum_{n = 1}^{N} {\left( \frac{O^{{\rm pred.}}_k\left(\alpha_i\right) - O^{{\rm exp.}}_k}{\sigma_k} \right)}^2 \, ,
\label{eq:chi2_def}
\end{align}
that is, the sum of the weighted difference squared between experimental data and model predicted values where the weight is the inverse of the experimental uncertainty. Usually a fit is deemed of good quality if the quantity $\chi^2_{\rm d.o.f.} \equiv \chi^2 / (N - N_{par})$, where $N_{\rm par.}$ is the number of parameters to be fitted, is close to one. Throughout the paper the parameter errors represent the 68 percent confidence interval for the parameter estimates. 

In equation (\ref{eq:chi2_def}) $O_k$ represents a generic data point of one or several observables mentioned in the previous sections. For the proton structure functions $\sigma_k$ is simply the experimental error of each point.
For total cross-section data we also need to take into account that some data points have uncertainties in the values of $s$ (e.g. in $\gamma \gamma \rightarrow X$ that is always the case because it is a measured quantity). To account for this we compute the total cross-section for $s + \Delta s$ and $s - \Delta s$, and compute
\begin{align}
\sigma_{\rm eff} = {\rm max} &\left(|\sigma^{\rm pred.}\left(s + \Delta s\right)-\sigma^{\rm pred.}\left(s\right)|\,, \right. \notag \\
& \left. |\sigma^{\rm pred.}\left(s - \Delta s\right)-\sigma^{\rm pred.}\left(s\right)| \right).
\end{align}
For these cases  $\sigma_k = \sqrt{{\left(\sigma_{\rm exp.}\right)}^2+{\left( \sigma_{\rm eff.}\right)} ^2}$ where $\sigma_{\rm exp.}$ is the experimental error.

For $\sigma\left(\gamma \gamma \rightarrow X\right)$ and  $\sigma\left(\gamma p \rightarrow X\right)$ we use the hadronic cross-section data files from the Particle Data Group~\cite{pdg_2018}. These data sets are a compilation of experimental results obtained in the last decades from several collaborations. The dataset of  $\sigma\left(\gamma p \rightarrow X\right)$ has cross-section values as a function of the laboratory momentum of the incoming on-shell photon. Hence we computed the respective center of mass energy $\sqrt{s}$ before performing the fits. We also considered only subsets of data with $\sqrt{s} > 4 \, {\rm GeV}$ for $\sigma\left(\gamma \gamma \rightarrow X\right)$ and  $\sigma\left(\gamma p \rightarrow X\right)$, yielding 39 and  45 experimental points respectively. The lower bound cuts result from the fact that our model does not realise the meson trajectory with an intercept around $0.35-0.55$. This trajectory dominates for smaller values of $s$ and the best our model can do is to reproduce data  in the intermediate range of $s$ through the third and fourth trajectories. In the future we plan to study how to include the meson trajectory and redo these fits. For the source of data for the proton structure functions $F_2\left(x, Q^2\right)$ and $F_L\left(x, Q^2\right)$ we use the HERA measurements of these observables available in~\cite{Aaron:2009aa, Collaboration:2010ry}. There are 249 points in the kinematical region with $x < 10^{-2}$ and $Q^2 \leq 400 {\rm \, GeV}^2$ for $F_2\big(x, Q^2\big)$, and 64 points with $x < 10^{-2}$ and $Q^2 \leq 45 {\rm \, GeV}^2$ for $F_L\big(x, Q^2\big)$. For the photon structure function 
$F_2^\gamma\big(x, Q^2\big)$ we consider the measurements of ALEPH~\cite{Barate:1999qy,Heister:2003an}, L3~\cite{Acciarri:1998ig} and 
OPAL~\cite{Ackerstaff:1997ng, Abbiendi:2000cw, Abbiendi:2002te} collaborations at LEP and of the TPC/Two Gamma collaboration \cite{Aihara:1986xw} at 
SLAC $e^+e^-$ storage ring PEP. These measurements contribute with 22 points with $x \leq 0.0235$ and $Q^2 \leq 17.8 {\rm \,GeV}^2$.

\begin{figure}[b!]
\center
\includegraphics[scale = 0.48]{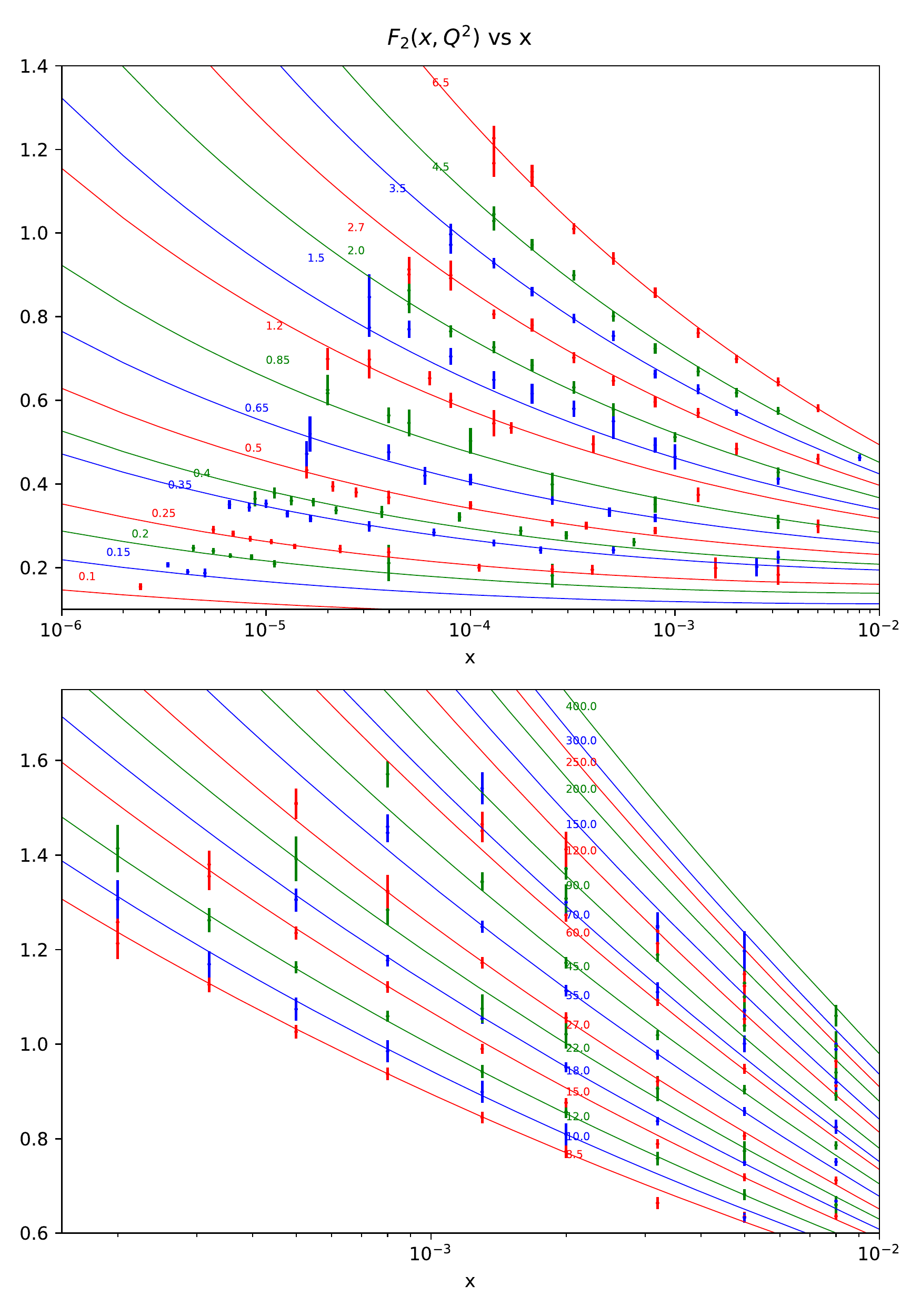} 
\caption{Predicted proton structure function $F_2\left(x,Q^2\right)$ vs experimental points. The  curves were obtained using  the values of table~\ref{table:GammaP_best_fit}.}
\label{fig:F2_best_fit}
\end{figure}

\begin{figure}[!t]
\center
\includegraphics[scale = 0.48]{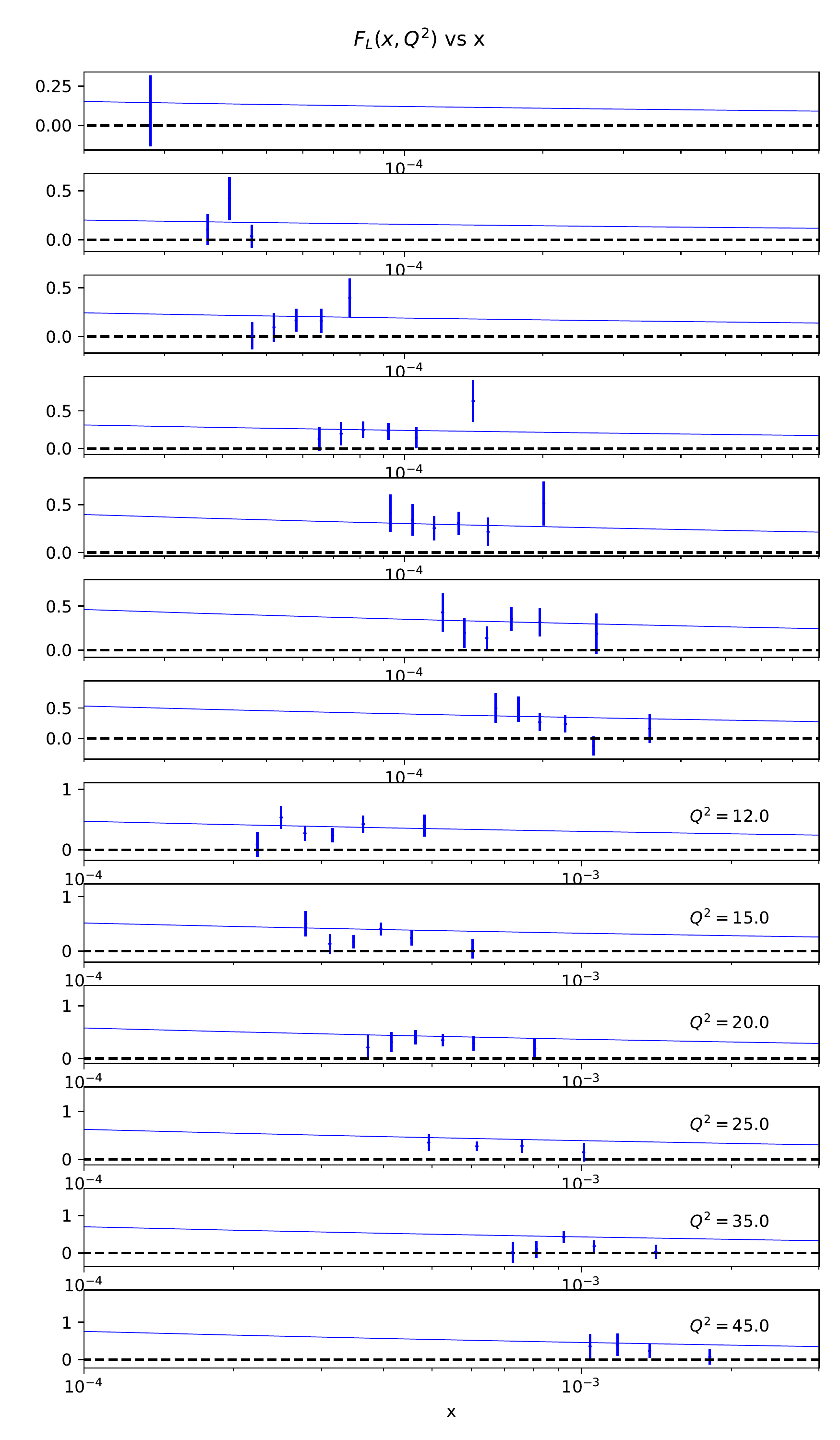} 
\caption{Predicted proton structure function $F_L\left(x,Q^2\right)$ vs experimental points. The curves were obtained using the values of table~\ref{table:GammaP_best_fit}.}
\label{fig:FL_best_fit}
\end{figure}

\begin{figure}[!h!]
\center
\includegraphics[scale = 0.47]{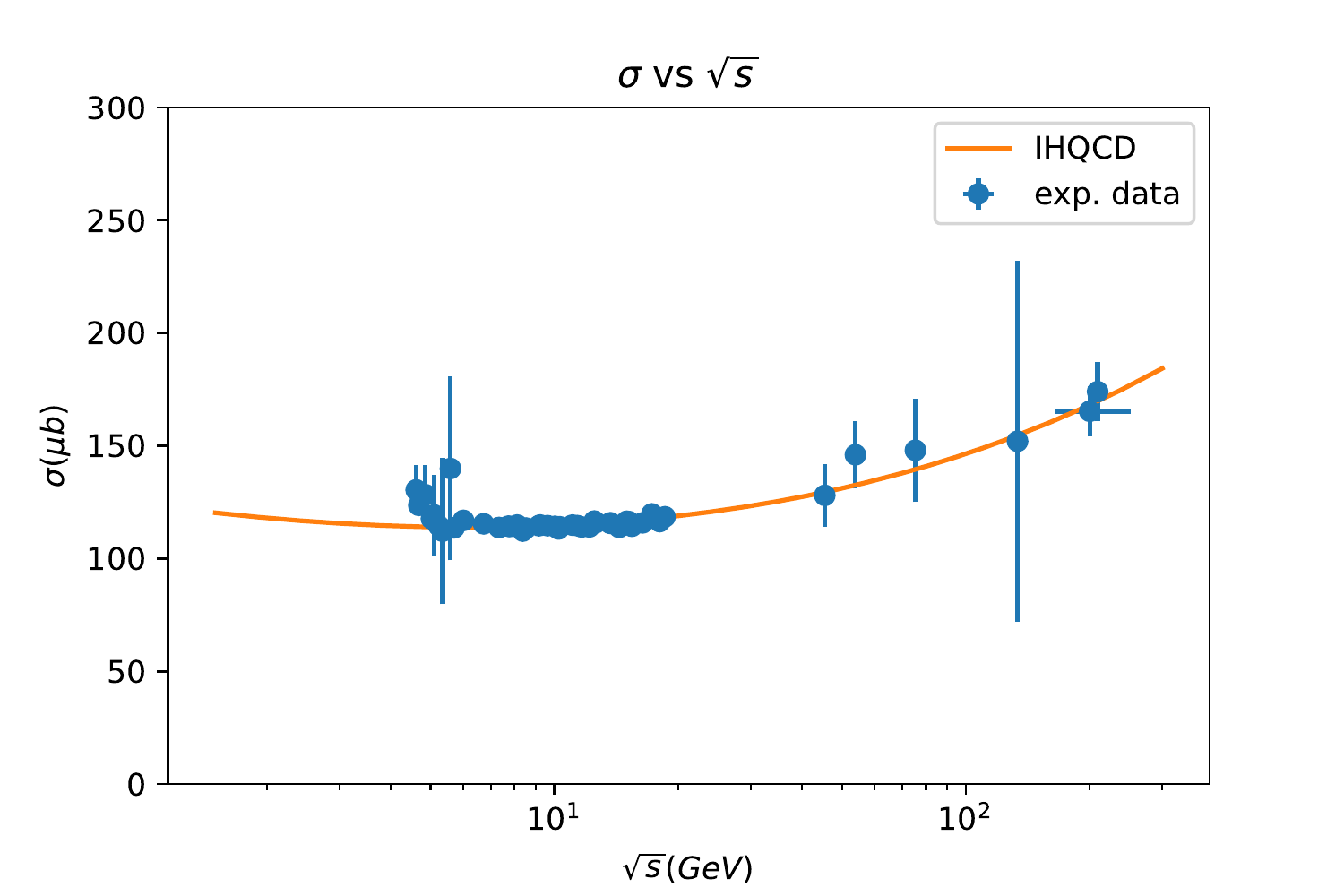} 
\caption{Predicted $\sigma \left( \gamma p \rightarrow X \right)$ vs experimental points. The curve was obtained using the values from table ~\ref{table:GammaP_best_fit}.}
\label{fig:SigmaGammaProton_best_fit}
\end{figure}

We now present the results of three fits we have performed. First we included only the data from observables of $\gamma^* p$ processes.
Using  data from the proton structure functions and of the total cross-section $\sigma(\gamma p \rightarrow X)$ the best fit values for the Pomeron kernel parameters and  constants ${\rm Im} \, g_n$ 
defined in  (\ref{eq:gn_def_gammap}) are present in table \ref{table:GammaP_best_fit}.  The corresponding intercepts of the Reggeons are also displayed in the same table.
The total number of experimental points used in this fit is 358 and a  $\chi^2_{d.o.f.}$ per degree of freedom of 1.40 was obtained. 
With those parameter values we compare the predictions of our model against the experimental data in figures \ref{fig:F2_best_fit}, \ref{fig:FL_best_fit} and  \ref{fig:SigmaGammaProton_best_fit}.
\begin{table}[t!]
\centering
\caption{Values of the parameters for the best joint fit of the proton structure functions $F_2\left(x, Q^2\right)$ and $F_L\left(x, Q^2\right)$ and $\sigma\left(\gamma p \rightarrow X\right)$ data from PDG with $x \leq 0.01$, $Q^2 \leq 400 \, \text{GeV}^2$ and $\sqrt{s} > 4.6 \, \text{GeV}$. We obtained   a $\chi^2_{d.o.f.}$ of 1.40 with 358 experimental points.}
\vspace{0.5cm}
\begin{tabular}{|c|c|c|}
\hline
Kernel parameters & couplings  & intercept \\
\hline
$a = -4.70 \pm 0.04$ & ${\rm Im} \, g_0 = -0.051 \pm 0.001$ & $j_0 = 1.165$\\ 
\hline
$b = 1.13 \pm 0.05$ & ${\rm Im} \, g_1 = 0.017 \pm 0.007$ & $j_1 = 1.082$ \\ 
\hline
$c =0.66 \pm 0.01$ & ${\rm Im} \, g_2 = -0.07 \pm 0.01$ & $j_2 = 0.977$ \\
\hline
$d = -0.098 \pm 0.006$ & ${\rm Im} \, g_3 = 0.358 \pm 0.007$ & $j_3 = 0.918$ \\ 
\hline
$l_s^{-1} = 6.47 \pm 0.08$ & & \\
\hline 
\end{tabular}
\label{table:GammaP_best_fit}
\end{table}

\begin{table}[t!]
\centering
\caption{Parameters for the fit with the photon structure functions $F_2^\gamma\left(x, Q^2\right)$ and $\sigma\left(\gamma \gamma \rightarrow X\right)$ data from PDG with $\sqrt{s} > 4 \, \text{GeV}$.  $\chi^2_{d.o.f.}=  1.07$ with 61 experimental points.}
\vspace{0.5cm}
\begin{tabular}{|c|c|}
\hline
couplings   & ${\rm value} \times 10^{-4} $\\
\hline
${\rm Im} \, g_0$  & $-1.789 \pm 0.065$\\ 
\hline
${\rm Im} \, g_1$  & $1.94 \pm 0.33$ \\ 
\hline
${\rm Im} \, g_2$  & $-3.19 \pm 1.11$ \\
\hline
${\rm Im} \, g_3$  & $13.37 \pm 1.21$ \\ 
\hline
\end{tabular}
\label{table:GammaGamma_best_fit}
\end{table}

\begin{figure}[!h]
\center
\includegraphics[scale = 0.42]{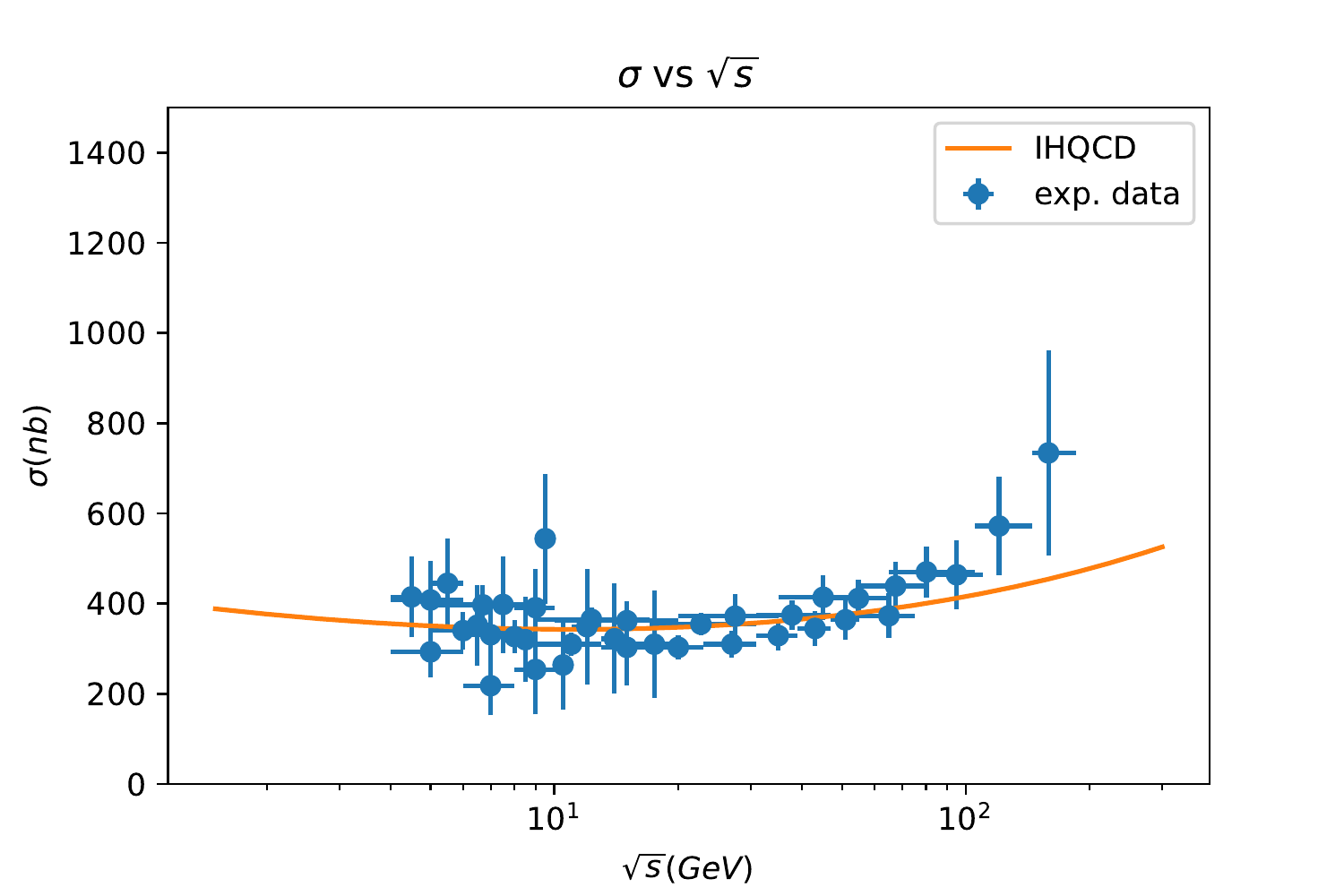} 
\caption{Predicted $\sigma\left(\gamma \gamma \rightarrow X\right)$ vs experimental points. The curve was obtained using the parameter values in table ~\ref{table:GammaGamma_best_fit}.}
\label{fig:SigmaGammaGamma_best_fit}
\end{figure}

\begin{figure}[!h]
\center
\includegraphics[scale = 0.43]{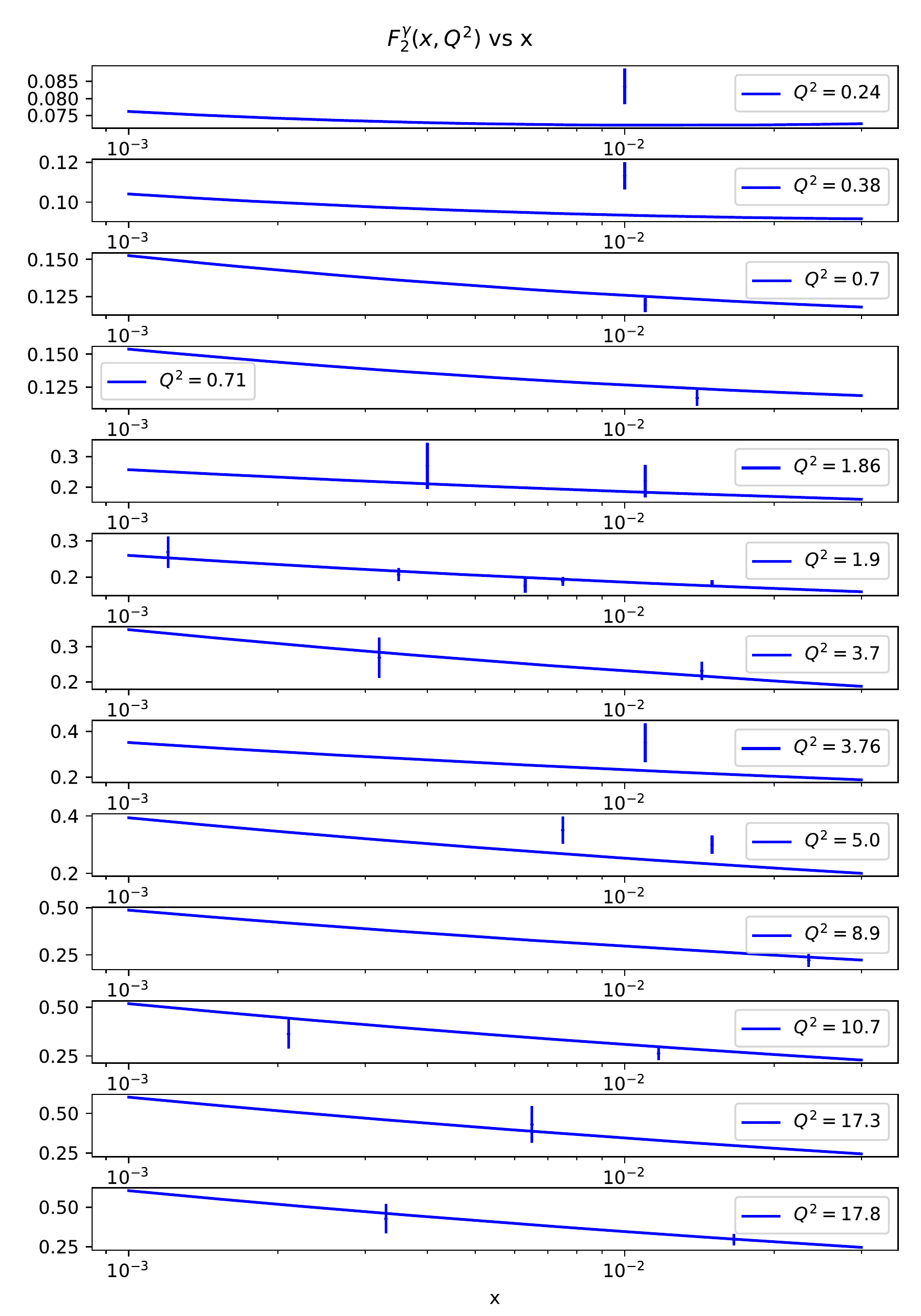} 
\caption{Predicted $F_2^\gamma\left(x, Q^2\right)$ vs experimental points. The curves were obtained using the values from table ~\ref{table:GammaGamma_best_fit}.}
\label{fig:F2Photon_best_fit}
\end{figure}

For the fit including the photon structure functions $F_2$ and the total cross section of $\gamma \gamma$ scattering, we obtained the values present in table~\ref{table:GammaGamma_best_fit}. We emphasize that in this fit the Pomeron kernel parameters were kept fixed and equal to the values in table \ref{table:GammaP_best_fit}. The number of points used in this fit was 61 and a $\chi^2_{d.o.f.}$ per degree of freedom of 1.07 has been obtained. The predictions of our model for these parameters against the $\sigma\left(\gamma \gamma \rightarrow X\right)$ and $F_2^\gamma$ data are displayed in figures~\ref{fig:SigmaGammaGamma_best_fit} and \ref{fig:F2Photon_best_fit}, respectively.

The good quality of the fits obtained so far and the definitions of ${\rm Im} \, g_n$ for $\gamma^* \gamma$ and of ${\rm Im} \, g_n$  for  $\gamma^* p$ processes suggest that a joint fit using all experimental data is possible. The fitting parameters in this fit are the gravitational couplings $\kappa_{j_n}$ between the bulk $U\left(1\right)$ gauge field with the $n$-th reggeon defined in (\ref{eq:gauge_field_spin_J_coupling}), and the product of the gravitational couplings $\bar{\kappa}_{j_n}$ of the bulk field dual to the proton in (\ref{eq:scalar_field_spin_J_coupling}) together with the corresponding $\bar{z}$ integral.
We have done such a fit making use of the combined 419 experimental points of the different observables. A $\chi_{d.o.f.}^2$ per degree of freedom of $1.38$ was obtained with the parameters of table~\ref{table:GammaGamma_GammaProton_best_fit}. For completeness the corresponding values of ${\rm Im} \, g_n$ for the $\gamma^* \gamma$ and $\gamma^* p$ processes are 
also given in table~\ref{table:GammaGamma_GammaProton_best_fit_2} and are very close to those found previously in tables~\ref{table:GammaP_best_fit} and \ref{table:GammaGamma_best_fit}.

\begin{table}[h!]
\centering
\caption{Values for the joint fit of $\gamma^*\gamma$ and $\gamma^*p$ processes. There are 419 experimental points giving a  $\chi^2_{d.o.f.}=1.38$.}
\vspace{0.5cm}
\begin{tabular}{|c|c|c|}
\hline
 \ \ $n$\ \ & $\kappa_{j_n}$ & $\bar{\kappa}_{j_n} \times \rm{\bar{z} \, integral }$ \\
\hline
$1$ & $0.05106 \pm 0.00093$  & $-18.040 \pm 0.335$  \\
\hline
$2$ & $0.0631 \pm 0.0055$ & $-5.45269 \pm 0.610$ \\
\hline
$3$ & $0.228 \pm 0.040$  & $10.0 \pm 1.8$ \\
\hline
$4$ & $-0.261 \pm 0.012$ & $-56.1 \pm 2.7$ \\
\hline
\end{tabular}
\label{table:GammaGamma_GammaProton_best_fit}
\end{table}
\begin{table}[h!]
\centering
\caption{Values of  ${\rm Im} \, g_n^{\gamma^*\gamma}$ and  ${\rm Im} \, g_n^{\gamma^*p}$ from the best fit parameters of table~\ref{table:GammaGamma_GammaProton_best_fit}.}
\vspace{0.5cm}
\begin{tabular}{|c|c|c|}
\hline
\ \ $n$\  \  &  ${\rm Im} \, g_n^{\gamma^*\gamma}$ &  ${\rm Im} \, g_n^{\gamma^*p}$ \\
\hline
$1$ & $-1.78883 \times 10^{-4}$  & -0.0510176\\
\hline
$2$ &  $1.93696 \times 10^{-4}$ &0.017369\\
\hline
$3$ &  $-3.19052 \times 10^{-4}$ & $-0.0744977$\\
\hline
$4$ &  $13.3664 \times 10^{-4}$ &0.357739 \\
\hline
\end{tabular}
\label{table:GammaGamma_GammaProton_best_fit_2}
\end{table}

Finally, we have also performed a global fit where the pomeron kernel parameters were also allowed to vary. We have found a $\chi^2_{d.o.f.}$ of 1.36 and parameter values close to the ones of 
tables~\ref{table:GammaP_best_fit},~\ref{table:GammaGamma_best_fit}  and \ref{table:GammaGamma_GammaProton_best_fit} therefore showing the consistency of our results.

%%%%%%%%%%%%%%%%%%%%%%%%%%%%%%%%%%%%%%%%%%%%%
\section{Gluon Parton distribution functions}
%%%%%%%%%%%%%%%%%%%%%%%%%%%%%%%%%%%%%%%%%%%%%

The structure functions $F_2$ and $F_L$ of the proton can be written in terms of the proton's PDFs. Their physical meaning is the probability density for finding a parton with a certain longitudinal momentum fraction $x$ at resolution scale $Q^2$. In the naive quark parton model $F_2= \sum_i e_i^2 x q_i (x)$, where the phenomena of Bjorken scaling is predicted, i.e. $F_2$ dependes only on $x$ and not $Q^2$.
The parton model also predicts that
\begin{equation}
F_L\big(x, Q^2\big) = F_2 \big(x, Q^2\big) - 2 x F_1\big(x, Q^2\big) = 0\,,
\end{equation} 
which is known as the Callan-Gross relationship and is satisfied if the partons inside the proton have spin-$\frac{1}{2}$.
These relations follow from considering only the QED diagram $\gamma^{*}$-parton and assuming that the partons have zero transverse momentum. At NLO QCD gluon radiation and $g\rightarrow q \bar{q}$ processes give rise to $\ln Q^2$ scaling violations and to partons with non-zero transverse momentum. Hence the Callan-Gross relation is no longer true, and $F_L$ can be related to $F_2$ and the gluon PDF $g(x, Q^2)$ through \cite{Altarelli:1978tq}
\begin{equation}
F_L\big(x,Q^2\big) = \frac{\alpha_s(Q^2)}{4\pi} \left( \frac{16}{3} I_F + 8 \bar{e}^2 I_G \right), 
\label{eq:F_L}
\end{equation}
where
\begin{align}
&I_F = \int_x^1 \frac{dy}{y} {\left(\frac{x}{y}\right)}^2 F_2 \big(y, Q^2 \big), \\
&I_G = \int_x^1 \frac{dy}{y} {\left(\frac{x}{y}\right)}^2 \left(1-\frac{x}{y}\right) \mathcal{G} \big(y, Q^2 \big),
\end{align}
and where $\mathcal{G}\big(y, Q^2\big) = y g(y, Q^2)$, $\alpha_s$ is the QCD coupling and $\bar{e}^2$ is the sum of the squares of the electric charges of the active quark flavours. The number of active flavours $n_f$ changes 
with the scale at which we want to evaluate the PDF. In this work we will assume that~$n_f = 3$ for~$Q^2 \leq m_c$, $n_f = 4$ for $m_c < Q^2 \leq m_b$ and $n_f = 5$ otherwise. $m_c$ and $m_b$ are the masses of the charm and bottom quarks, respectively. We wish to express the gluon PDF $g(x, Q^2)$ in terms of the structure functions $F_2$ and $F_L$. This can be done by computing the derivative  with respect to $x$ of (\ref{eq:F_L}) and using the definition of
the integrals $I_F $ and $I_G$.  A straightforward computation yields
\begin{align}
 \bar{e}^2 \mathcal{G} \big(x, Q^2 \big) =& \left( -2  + \frac{2x}{3} \frac{\partial }{\partial x}\right) F_2\big(x, Q^2 \big)+
 \nonumber \\
 &
 \frac{\pi}{\alpha_s(Q^2)} \left(   3 - 2x \frac{\partial }{\partial x} + \frac{x^2}{2}  \frac{\partial^2 }{\partial x^2}  \right)  F_L\big(x, Q^2 \big) \, .
 \label{eq:G(x)}
 \end{align}

We use the NLO result (\ref{eq:G(x)}) to estimate the function $\mathcal{G} \big(x, Q^2 \big)$ at small $x$ from the holographic values for $F_2$ and $F_L$ computed using the best fit parameters in table~\ref{table:GammaP_best_fit}.  We shall compare our gluon PDFs with the CT18 and NNPDF collaborations \cite{Hou:2019efy, Ball:2017nwa}, also at NLO. For the coupling constant  $\alpha_s(Q^2)$ in  (\ref{eq:G(x)})  we used the holographic 
value (see \cite{Ballon-Bayona:2015wra} for a plot of this function). 

\begin{figure}[h!]
  \center
  \includegraphics[scale = 0.4]{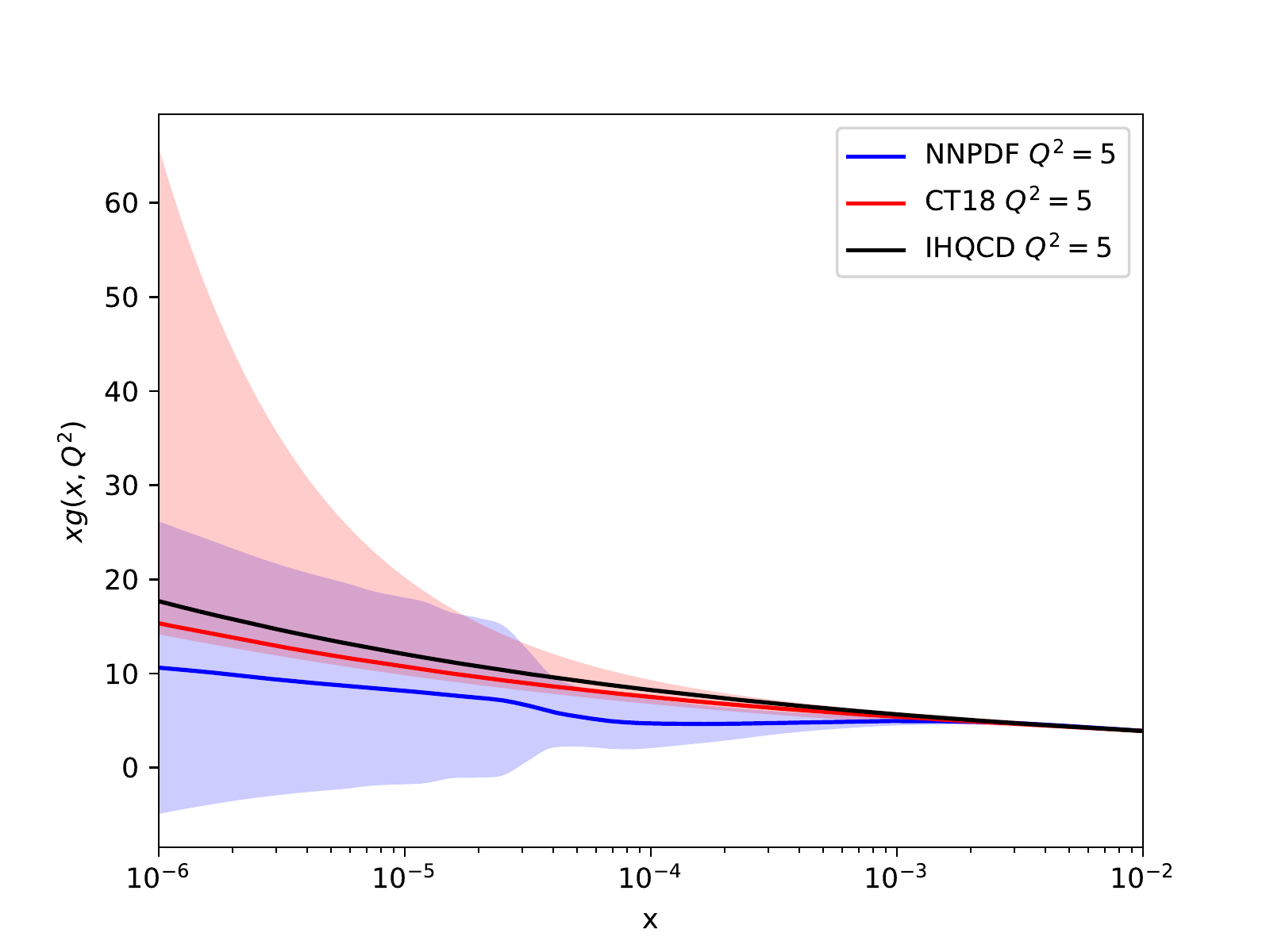} 
   \includegraphics[scale = 0.4]{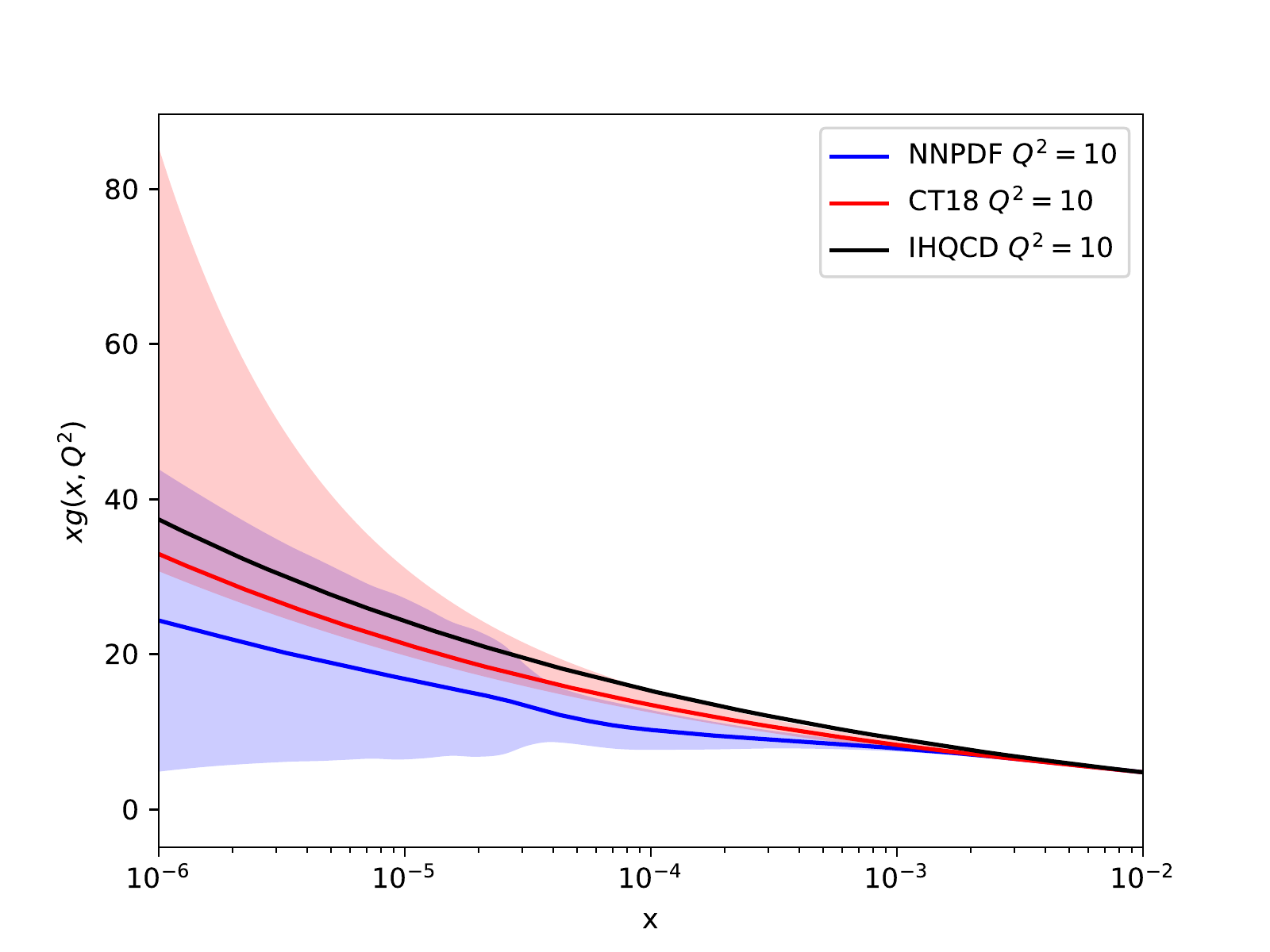} 
    \includegraphics[scale = 0.4]{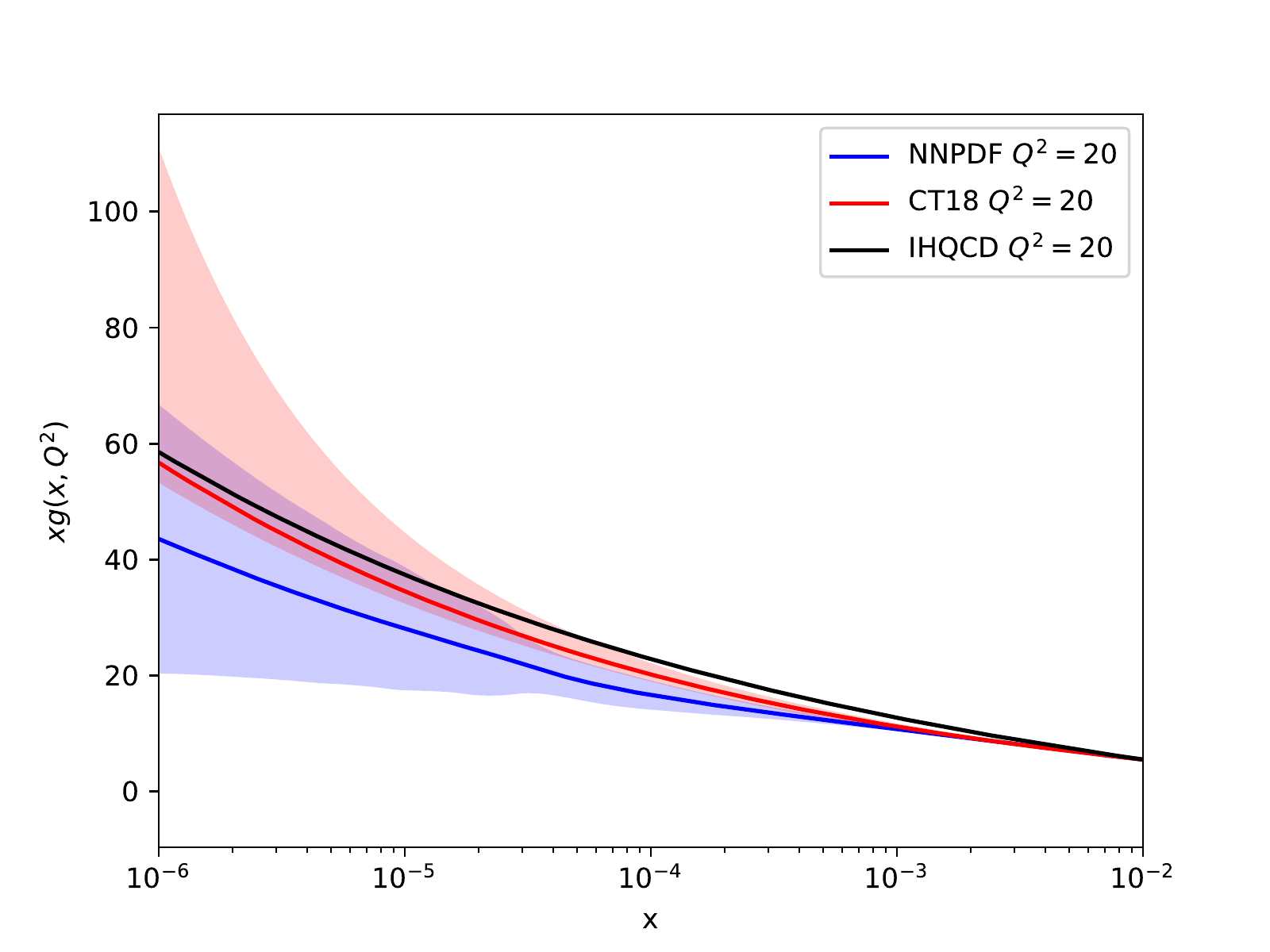} 
     \includegraphics[scale = 0.4]{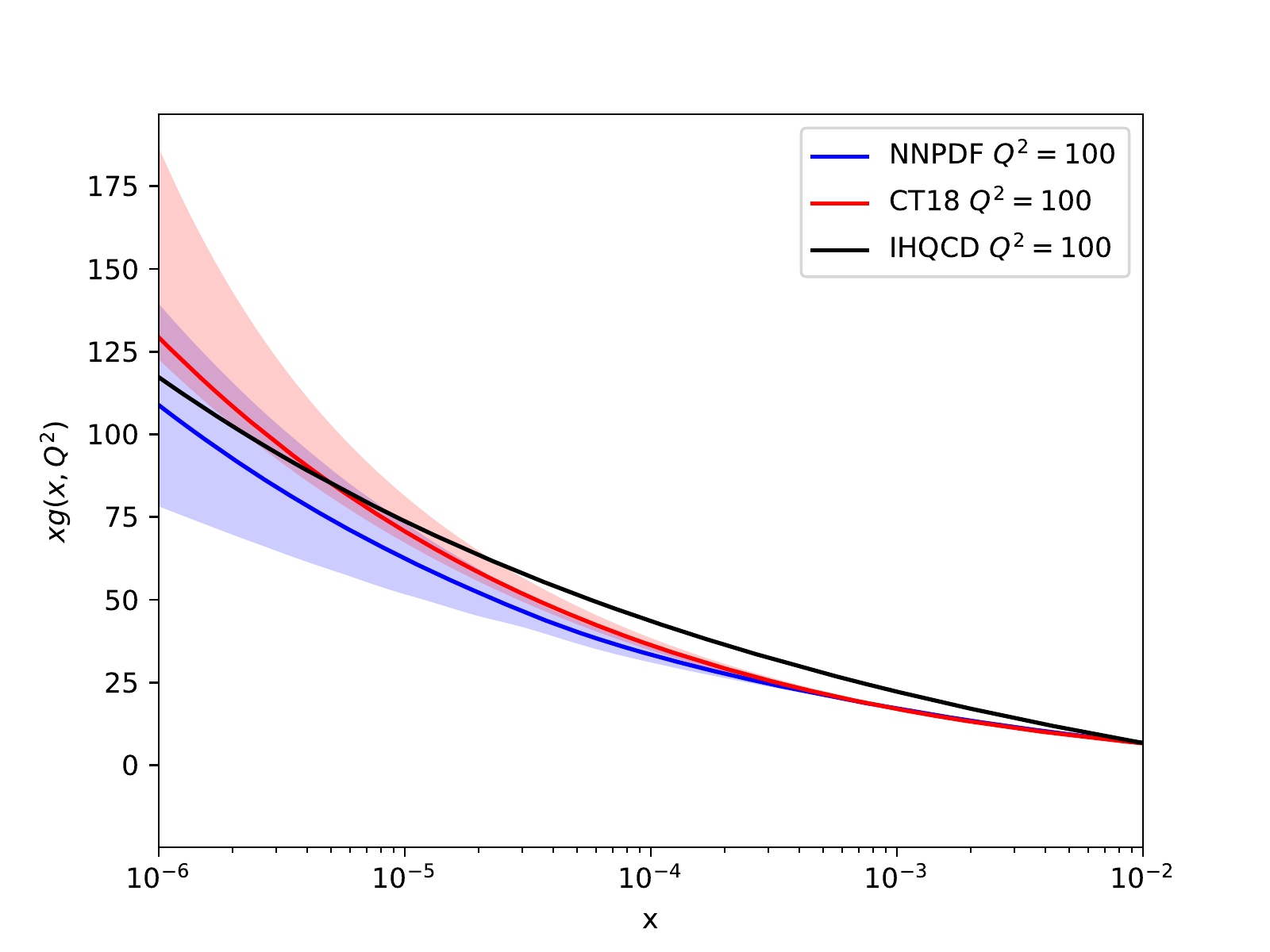} 
  \caption{Comparison between the holographic  gluon PDFs with the ones of CT18NNLO and NNPDF for $Q^2 = 5, \, 10, \, 20, \, 100\, {\rm GeV}^2$.}
\label{fig:gluonPDF}
\end{figure}

Since  the functions $F_2$ and $F_L$ do not reproduce the data in the range $10^{-2}<x<1$,  and in particular have the wrong asymptotics for $x\to1$,  we do not expect to be able to reproduce  $\mathcal{G} \big(x, Q^2 \big)$ in the transition region $x\sim10^{-2}$ where there are other contributions to the structure functions from quarks. To match the correct asymptotic values  in this transition region, we added a constant function $f(Q^2)$ to  $\mathcal{G} \big(x, Q^2 \big)$ such that we match the value of $\mathcal{G} \big(x, Q^2 \big)$ given by the average of the other collaborations at the specific value $x=10^{-2}$.  Our main goal is to assess whether or not we can predict the correct low $x$ evolution of the gluon PDFs starting from $x = 10^{-2}$ to lower values of $x$.
Our results are presented in figure~\ref{fig:gluonPDF}.  It clear that we are able to reproduce the correct behaviour within the other collaborations allowed regions. Of course these results should be taken as a simple qualitative indication, since we are using a NLO expression for the gluon PDFs together with the holographic structure functions and coupling constant. Also, as mentioned above, the resulting PDF's do not have the correct asymptotics for $x \rightarrow 1$ \cite{Brodsky:1989db, Brodsky:1994kg}, but this is expected since the whole analysis is only valid in the low $x$ region.

%%%%%%%%%%%%%%%%%%%%%%%%%%%
\section{Conclusion}
%%%%%%%%%%%%%%%%%%%%%%%%%%%
In this paper we extended the previous work of  \cite{ballon_bayona_unity_2017}, that considered the improved holographic QCD model to study the proton structure function $F_2$ at low $x$, to include the proton longitudinal structure function $F_L$, the total cross-section $\sigma(\gamma p \rightarrow X)$, the photon structure function $F_2^\gamma$ and the total cross-section of $\gamma \gamma$ scattering.
The $\chi^2$ quality of our fit improves on this previous work from $\chi^2=1.7$ to $\chi^2=1.4$. This is due to the fact that, in addition to the $249$ data points  from the structure function $F_2$, 
we have an extra of $64$ data points from the structure function $F_L$ and $45$ data points from the cross-section $\sigma(\gamma p \rightarrow X)$, which can also be described  holographically. 
These results are obtained in a very large kinematical window of $x < 10^{-2}$ and $Q^2 \le 400\  {\rm GeV}^2$ for $F_2$,
$Q^2 \le 45\  {\rm GeV}^2$ for $F_L$ and $\sqrt{s} > 4.6 \, {\rm GeV}$ for $\sigma(\gamma p \rightarrow X)$.

Then we have shown that the Pomeron kernel found in the fit of the $\gamma^{*} p$ processes could be used to achieve excellent fits of the $\gamma^{*} \gamma$ processes discussed in section~\ref{sec:gamma_gamma}. We have also checked from a global fit, where we include all the processes and vary all the parameters, that the fitting parameters do not vary much, showing the consistency of the model. To our knowledge, this is the first text of the IHQCD pomeron model in such a  vast class of processes in a wide kinematical range.

Using the 
NLO relation (\ref{eq:G(x)}) we were  able to reproduce the low $x$ evolution of gluonic PDFs using as input  the holographic functions $F_2$, $F_L$  and $\alpha_s(Q^2)$.
In the region of larger $Q^2$ there is  more tension in matching to the PDFs of the other collaborations, as can be seen in the $Q^2=100\ {\rm GeV}^2$ plot of 
figure \ref{fig:gluonPDF}.  
According to equation  (\ref{eq:G(x)}), in this region it is essential to have a good description of $F_L$ because it is divided by $\alpha_s$ which is small for high values of $Q^2$ due to asymptotic freedom.
Since we are performing a $\chi^2$ fit to the data, the fitting process favours a good description of $F_2$ because the uncertainties are lower than those of $F_L$ as compared with the value measured. The uncertainties of $F_L$ are of the same size as the measured value.  Thus, better measurements of $F_L$ might help our model give a better description of the gluon PDFs.
Moreover, for high values of $Q^2$ it is important to include heavy quarks in global QCD fits. Nowadays PDF groups use variable flavour schemes in order to produce high quality results. To holography this means that the holographic dual must contain quark flavour degrees of freedom if the model ought to be successful at computing them. The IHQCD model we used as our QCD vacuum exhibits the properties of large $N_c$ Yang-Mills, including the running of the coupling constant.
Following the  ideas of \cite{Jarvinen:2011qe}, it would be very interesting to include flavour degrees of freedom in this holographic QCD model and to test if the quality of our fits generically
improve, mainly for high $Q^2$. Including  quarks is actually necessary because it is believed that the third and fourth dominant Regge trajectory actually come from the mesonic sector, instead of the glueball sector.

Another problem would be to determine holographically the gluonic PDFs, without making reference to perturbative QCD definitions. Holography may actually be the right set up for a non-perturbative 
definition. In fact, gluonic PDFs can be defined using a Wilson loop operator  (see for example \cite{Collins:1981uw}). Thus, it would be very interesting to use the duality between Wilson loops and strings in the dual
geometry to explore this problem.

\section{Acknowledgments}

This research received funding from the Simons Foundation grants 488637  (Simons collaboration on the Non-perturbative bootstrap)
and from the  grant CERN/FIS-PAR/0019/2017. 
Centro de F\'\i sica do Porto is partially funded by Funda\c c\~ao para a Ci\^encia e a Tecnologia (FCT) under the grant
UID-04650-FCUP.
 AA is funded by FCT under the IDPASC doctorate programme with the fellowship  PD/BD/114158/2016.

\bibliographystyle{elsarticle-num}
\bibliography{photonScattering_notes}

\end{document}